\documentclass[lettersize,journal]{IEEEtran}
\usepackage{amsmath,amsfonts,amssymb}
\usepackage[ruled,linesnumbered]{algorithm2e}
\usepackage{array}
\usepackage[caption=false,font=normalsize,labelfont=sf,textfont=sf]{subfig}
\usepackage{textcomp}
\usepackage{stfloats}
\usepackage{url}
\usepackage{verbatim}
\usepackage{graphicx}
\usepackage{cite}
\usepackage{svg}
\usepackage{color}
\usepackage{orcidlink}
\hypersetup{hidelinks}	
\usepackage{enumitem} 
\usepackage{booktabs}
\usepackage{multirow}
\usepackage{array}      
\usepackage{ragged2e}   
\usepackage{microtype}  
\usepackage{diagbox}    
\usepackage{siunitx}    
\usepackage{etoolbox}   
\usepackage{adjustbox}
\usepackage{tabularx}
\usepackage{svg}
\usepackage{float}
\usepackage{hyperref}
\usepackage[capitalize]{cleveref}
\usepackage[switch]{lineno} 
\usepackage{tikz}
\usetikzlibrary{shadows, calc} 

\hypersetup{
	colorlinks=true,
	linkcolor=blue,
	citecolor=blue,
	urlcolor=blue
}

\crefformat{equation}{\textcolor{blue}{(#2#1#3)}}
\Crefformat{figure}{\textcolor{blue}{(#2#1#3)}}
\Crefformat{cite}{\textcolor{blue}{(#2#1#3)}}

\begin{document}

	\title{Approximate Gaussian Mapping for Generative Image Steganography}
	
	\author{Yuhua Xu$^{\orcidlink{0009-0007-6624-8681}}$, Wei Sun, Jiaxing Lu, Jingying Zhou, Chen Gu,  Chengpei Tang$^{\orcidlink{0000-0002-8139-6742}}$
		\thanks{Corresponding author: Wei Sun and Chengpei Tang.}
		\thanks{Yuhua Xu, Wei Sun, Jiaxing Lu, Jingying Zhou and Chen Gu are with the School of Electronics and Information Technology (School of Microelectronics), Sun Yat-Sen University, Guangzhou 510006, China. e-mail: sunwei@mail.sysu.edu.cn; \{xuyh55, lujx63, zhoujy229, guch9\}@mail2.sysu.edu.cn}
		\thanks{Chengpei Tang is with the School of Advanced Manufacturing, Sun Yat-Sen University, Shenzhen 518107, China. e-mail: tchengp@mail.sysu.edu.cn}
	}
	
	\markboth{Journal of \LaTeX\ Class Files,~Vol.~14, No.~8, August~2021}%
	{Shell \MakeLowercase{\textit{et al.}}: A Sample Article Using IEEEtran.cls for IEEE Journals}
	
	\IEEEpubid{0000--0000/00\$00.00~\copyright~2021 IEEE}
	
	\maketitle

\begin{abstract}
	Ordinary differential equation (ODE)-based diffusion models enable deterministic image synthesis, establishing a reversible mapping suitable for generative steganography.
	While prevailing methods strictly adhere to a standard normal prior, empirical evidence indicates that controlled deviations from this distribution reduce numerical inversion errors without compromising perceptual quality.
	Leveraging this observation, the Approximate Gaussian Mapping (AGM) is proposed as a linear transformation strategy that embeds secrets by modulating noise scale and variance.
	To balance retrieval numerical consistence and security, a two-stage decoupled optimization strategy is introduced to minimize the Kullback-Leibler divergence subject to target bit accuracy constraints.
	Beyond the proposed method, we conduct a mechanistic analysis of the divergent behaviors between pixel-space and latent-space architectures.
	The experimental results reveal that the VAE encoder enhances robustness by filtering external perturbations, 
	whereas the structural regularization of the VAE decoder and the semantic variance introduced by text prompts jointly mask embedding artifacts to improve security.
	Experimental results confirm that pixel-space implementations maximize embedding capacity for lossless channels, while latent-space approaches offer superior robustness and security suitable for adversarial environments.
\end{abstract}

	
\section{Introduction}
\label{sec:introduction}
\IEEEPARstart{T}{he} rise of generative AI is not only transforming content creation but also altering the technical approaches of image steganography.
For decades, research in this field has focused on minimizing embedding distortion through advanced cost design \cite{HUGO, S-UNIWARD, HILL, WOW}, with Syndrome-Trellis Codes (STC) providing a near-optimal framework under these constraints \cite{STC}.
However, this strategy faces an inherent limitation that even a small form of modification would leave statistical artifacts that can be detected by modern deep learning-based steganalysis \cite{SRNet, XuNet, YeNet, SiaStegNet}.
The development of generative models \cite{DDPM, GAN, Glow} offer a direct way to overcome this limitation. It shifts the objective from imperceptible alteration to generation.
A stego image produced from secret data is no longer a degraded version of an existing one but a sample statistically consistent with other generated images. 
This property ensures statistical indistinguishability from benign samples, as no original reference exists for comparison.

The transition from modification-based embedding to data origination is now realized through Generative Image Steganography (GIS). 
In this context, diffusion models \cite{Score_Function} offer a particularly suitable structure, 
characterized by a probability flow ODE that couples the noise $\mathbf{x}_T$ sampled from standard normal prior (SNP) and the high-fidelity image $\mathbf{x}_0$ as the terminal and initial boundary conditions of a unique integral trajectory.
This mathematical formulation guarantees a deterministic and reversible mapping, thereby providing a direct mechanism for GIS tasks.
Although this reversibility is theoretically ensured by the underlying ODE formulation, practical implementations rely on numerical solvers that introduce minor approximation errors. 
Recent advances such as denoising diffusion implicit model (DDIM) \cite{DDIM} and the DPM-Solver family \cite{dpm_solver,dpm_solver_++} have substantially mitigated these errors, 
reducing the number of inference steps from thousands to tens while maintaining stable numerical precision. 
In practice, these improvements ensure that the bit accuracy rate (BAR) in information recovery remains consistently high, enabling reliable and efficient GIS operation.

Existing studies \cite{Stegaddpm, CLT, DCT_noise, Position} have leveraged this property by designing reversible mappings that convert secret data into noise statistically consistent with the model’s prior distribution. 
Although these approaches achieve high-fidelity generated image quality, they often suffer from substantial computational overhead or limited embedding capacity. 
These works rely on the assumption that strict conformity to the prior must be maintained for reliable GIS, without examining the model’s behavior under non-standard noise conditions. 
Furthermore, current research restricts experimentation to either pixel-space diffusion or latent-space diffusion \cite{LDM} based on variational autoencoder (VAE) \cite{VAE} and text prompts embedding model CLIP \cite{CLIP}, neglecting direct cross-space analysis. 
This omission leaves unresolved how architectural differences influence security, robustness, and embedding efficiency, despite the fact that such comparative evaluation is essential for understanding the intrinsic properties of GIS systems.
\IEEEpubidadjcol
\begin{figure*}[!t]
	\centering
	\includegraphics[width=\textwidth]{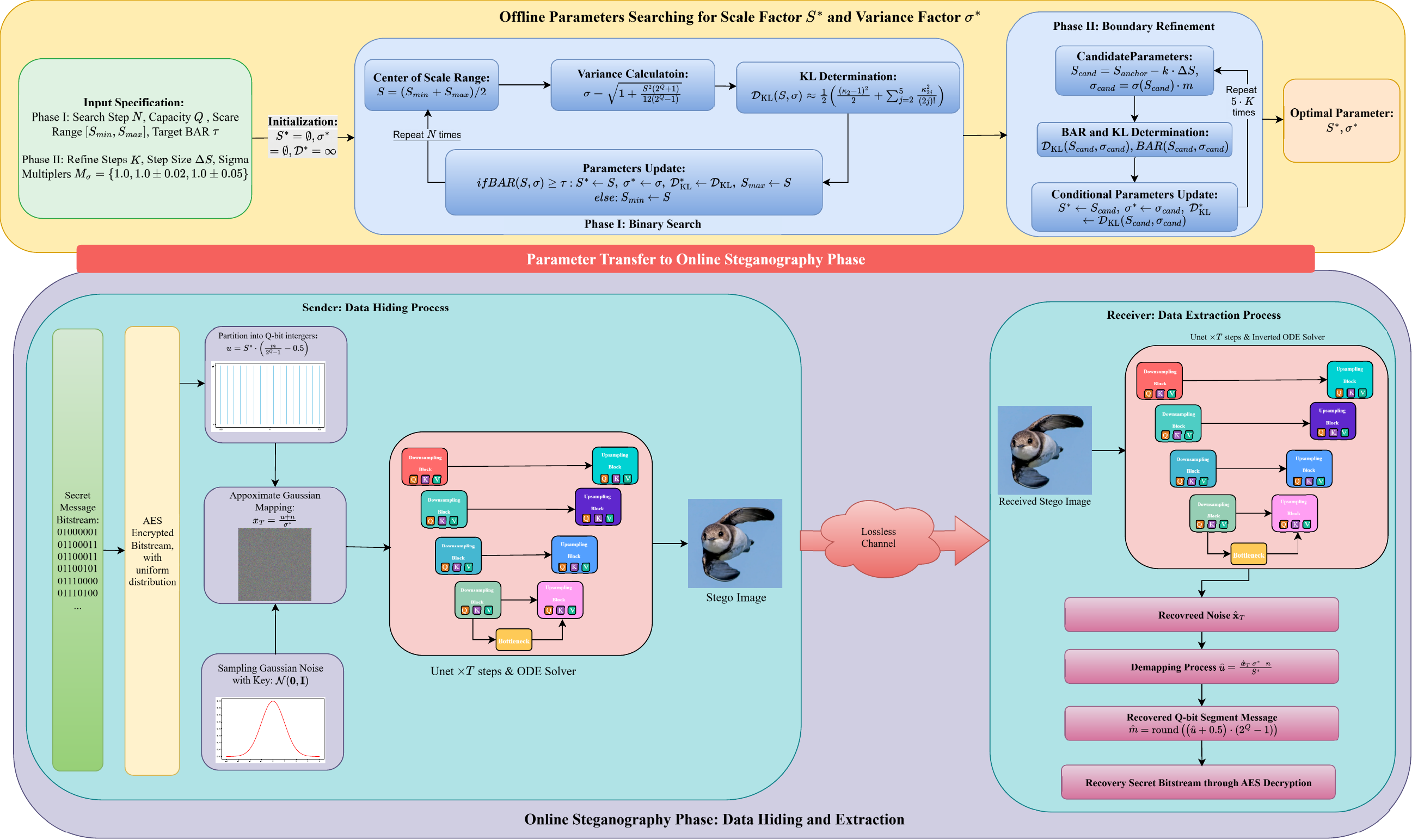} 
	\caption{Framework overview comprising offline parameter optimization and online steganography transmission. 
	The offline phase determines optimal mapping parameters ($S^*, \sigma^*$) by minimizing analytical Kullback-Leibler divergence subject to target recovery bit accuracy rate constraints. 
	The online phase utilizes these parameters within the Approximate Gaussian Mapping to transform encrypted secrets into noise $\mathbf{x}_T$, enabling reversible image synthesis and message retrieval via ODE integration.}
	\label{fig:overview}
\end{figure*}

To investigate the joint influence of model architecture and prior assumptions, we first examine how deviations from the prior affect the noise recovery behavior, thereby assessing the necessity of SNP conformity in diffusion-based steganography. 
The analysis reveals that strict adherence to the SNP is not a prerequisite for optimal recovery, instead, the controlled perturbations can improve numerical reversibility. 
Leveraging this observation, the Approximate Gaussian Mapping (AGM) is proposed as a computationally efficient linear transformation that relaxes strict SNP constraints to balance fidelity and embedding capacity.
Furthermore, a cross-space analysis across pixel and latent architectures elucidates the mechanisms governing their divergent behaviors, establishing distinct deployment criteria for differing security environments.
The main contributions can be summarized as follows: 
\begin{itemize}	
	\item The AGM strategy is introduced based on the empirical finding that a controlled deviation from the SNP can enhance the numerical reversibility of the diffusion ODE. 
	As a specific realization of approximate priors, this method incorporates a flexible optimization objective that enables negotiation between statistical imperceptibility and message recovery accuracy, 
	thereby adapting to diverse security and reliability constraints.
	\item This study elucidates the underlying mechanisms of robustness and better anti-steganalysis detection performance in latent diffusion steganography. 
	Experimental results indicate that the VAE encoder as a structural regularizer that attenuates high-frequency perturbations 
	and indicates that the high intrinsic variance introduced by semantic conditioning effectively masks embedding artifacts from detectors.
	\item A systematic comparative evaluation of diffusion-based steganography is conducted across both pixel and latent spatial domains. 
	The results establish a deployment criterion where pixel-space implementations maximize embedding capacity for secure lossless channels, 
	whereas latent-space approaches provide superior robustness and security suitable for transmission in adversarial or open-channel environments.
\end{itemize}

The remainder of this paper is organized as follows. 
\cref{sec:related_work} reviews pertinent research on steganography. 
\cref{sec:preliminaries} introduces the theoretical basis of diffusion models, including their ODE formulation and advanced numerical solvers.
\cref{sec:proposed_method} presents the proposed framework in detail, covering the noise mapping, data embedding and extraction procedures, as well as the two-stage parameters optimization strategy.
\cref{sec:experiments} describes the experimental setup, evaluation metrics, and provides a comprehensive analysis of the empirical results.
Finally, \cref{sec:conclusion} concludes the paper with a summary of key findings and a discussion of potential future research directions.

\section{Related Works}
\label{sec:related_work}
The evolution of image steganography has been characterized by a continuous pursuit of higher embedding capacity, improved imperceptibility, and greater resilience against steganalysis. 
This pursuit has driven the field through two distinct methodologies, from methods that modify existing cover images to approaches that synthesize stego images directly from secret messages.

\subsection{Steganography With Existing Cover Images}
Traditional steganography aims to embed secret data while introducing minimal modification to the cover image. 
Early techniques such as Least Significant Bit (LSB) substitution \cite{LSB} and its matching variants \cite{LSBM} exploit the insensitivity of human vision to small pixel variations. 
However, such changes introduce detectable artifacts in the cover histogram, allowing basic statistical tests such as the chi-square analysis \cite{chi2} to reveal the presence of hidden data. 
This limitation shows that visual imperceptibility alone is insufficient and motivates the development of adaptive steganography, which embeds data in complex or textured regions to better preserve the cover’s statistical properties.

This adaptive principle was formalized through the design of distortion learning functions, which assign modification costs to individual cover elements such as pixels or frequency domain coefficients. 
The embedding process is formulated as a constrained optimization problem that minimizes total distortion under the requirement of successful message embedding. 
Syndrome-Trellis Codes (STC) \cite{STC} provide an efficient framework for this formulation by decoupling coding implementation from distortion design. 
This separation allows focused research on the construction of more precise cost functions, leading to methods such as HUGO \cite{HUGO}, S-UNIWARD \cite{S-UNIWARD}, HILL \cite{HILL}, and WOW \cite{WOW}. 
Subsequent works \cite{ASDL-GAN,UT-GAN,SPAR-RL} extended this framework by introducing deep learning to infer the potential embedding costs, reducing dependence on handcrafted distortion models.
The use of deep learning for distortion learning naturally raises the question of whether the embedding process itself can also be learned, and this line of inquiry has motivated works such as \cite{Hidden,Hayes2017Generating, Zhang2019Invisible} that realize end-to-end steganography frameworks where both distortion modeling and message embedding are learned implicitly.
To achieve a tighter coupling between embedding and decoding, several studies \cite{Lu2021Large, Hinet} employed invertible neural networks (INNs), which model the steganography process as a bijective transformation to ensure reversible message extraction. 
Nevertheless, the modification-based nature of these approaches inevitably leaves subtle artifacts in the cover image. 
Recent advances in deep learning–based steganalysis, including SRNet \cite{SRNet}, XuNet \cite{XuNet}, YeNet \cite{YeNet}, and SiaStegNet \cite{SiaStegNet}, have demonstrated strong capability in detecting such artifacts by learning high-dimensional representations of embedding traces.

Some works \cite{SWE, DWT, recognition, classification} explored selection-based steganography, which encodes secret data by choosing cover images from large databases that match specific feature constraints. 
These methods do not modify the covers directly, but they are constrained by the need for extensive image collections, limited embedding capacity, and potential security vulnerabilities. 
These limitations have motivated the development of generative approaches that create stego images directly from secret information.

\subsection{Generative Image Steganography}
Generative Image Steganography (GIS) embeds secrets by synthesizing cover images directly from encrypted latent variables, thereby eliminating the dependence on pre-existing covers. 
The field has evolved through three main architectural paradigms: GANs, Normalizing Flows, and Diffusion Models.

\subsubsection{GAN-Based Steganography}
Early GAN-based methods produced plausible images first and then embedded secrets \cite{StegGAN, SSGAN}, while more recent approaches integrate secrets into the generative process, including semantic mapping \cite{syn_with_seman}, 
structural and textural mapping \cite{syn_with_structure}, or feature map modulation \cite{Stegastylegan_first}.
For instance, Su et al. \cite{Stegastylegan_second} attempted to embed data within StyleGAN2 by modulating the sign of noise elements to preserve the Gaussian prior $\mathcal{N}(\mathbf{0},\mathbf{I})$. 
However, this operation obliterates the original sign information, rendering unambiguous recovery impossible without side information. 
This limitation underscores the intrinsic architectural deficiency of GAN-based steganography that the non-invertible nature of the generator precludes analytical reversal from the image back to the noise domain. 
Consequently, such frameworks necessitate the training of auxiliary extractor networks \cite{First_GAN_Stega},  resulting in persistently suboptimal BAR.

\subsubsection{Flow-Based Steganography}
To achieve lossless recovery, subsequent works \cite{Position, GSF} adopted Invertible Neural Networks (INNs) or Normalizing Flows \cite{Glow}, which construct a bijective mapping between SNP noise and images.
S2IRT \cite{Position} encodes secrets via sample permutations within $[-2, 2]$, but the associated dynamic combinadic ranking involves large-number arithmetic that scales super-linearly, 
incurring computational latency comparable to the generative inference itself. 
GSF \cite{GSF} modifies IEEE 754 floating-point bits; while sign-bit modulation preserves symmetry, embedding into the mantissa for higher capacity severely distorts the continuous Gaussian density. 
GSF employs non-standard capacity metrics based on 32-bit precision rather than the 8-bit pixel format, yielding inflated embedding rates. 
More critically, despite theoretical invertibility, Flow-based models exhibit limited synthesis fidelity, which undermines the statistical consistency of generated images and exposes them to steganalysis detection.

\subsubsection{Diffusion-Based Steganography}
Diffusion Probabilistic Models (DPMs) have emerged as the dominant framework by integrating high-fidelity synthesis with mathematical reversibility via advanced ODE solvers \cite{dpm_solver, dpm_solver_++}. 
Although numerical integration introduces minor inversion errors, such recovery deviations remain within the tolerance defined by the code distance of the secrets, 
allowing diffusion-based steganography to maintain higher BAR than other generative architectures. 

Within this framework, steganography is formulated as the reverse diffusion process \cite{Stegaddpm,CLT,DCT_noise,Cross,SYSU_Lu}, where secret-infused Gaussian noise vectors serve as the initial states $\mathbf{x}_T$.
The transformation of secrets into SNP-compatible noise is critical in diffusion-based steganography, as most GIS implementations rely on open-source pre-trained models that assume SNP inputs.

Existing embedding strategies are categorized by their approach to this distribution matching problem. Exact distribution-preserving methods enforce strict mathematical correspondence to the SNP. 
StegaDDPM \cite{Stegaddpm} maps secrets into SNP noise via the inverse cumulative distribution function (ICDF). 
To achieve deterministic sampling within the inherently stochastic denoising diffusion probabilistic model (DDPM) framework \cite{DDPM}, the method fixes the random seeds used for noise sampling across all $T$ time steps.
Analogous to GSF \cite{GSF}, the reported embedding rate is also calculated based on 32-bit floating-point representations, resulting in numerically inflated capacity metrics.
Chen et al. \cite{SYSU_Lu} encode secret bits via rejection sampling from SNP, mapping 0 to the left tail and 1 to the right tail according to quantile-defined boundaries of embedding probability. 
The central region remains unselected, thus preserving the prior distribution. 

Conversely, strategies based on asymptotic convergence leverage limit theorems to approximate the SNP.
Hu et al. \cite{CLT} leverage the central limit theorem (CLT) to map secrets into Gaussian noise explicitly via a large random orthogonal matrix.
Zhou et al. \cite{DCT_noise} employ the inverse discrete cosine transform (IDCT), whose underlying operation can be interpreted as an orthogonal matrix transformation. 
By the CLT, the resulting noise approximates SNP after sufficient mixing.
Yu et al. \cite{Cross} transform secrets into Gaussian noise by leveraging the diffusion model's forward noising process under one prompt and subsequently generate an unrelated image under a second prompt. 
However, if the two prompts differ substantially, the reconstructed secret after the full encode-decode cycle can deviate significantly from the original, reducing recovery accuracy.

\section{Preliminaries}
\label{sec:preliminaries}
This section presents diffusion models with deterministic reformulation and numerical solvers facilitating efficient and reversible generation for steganography use.
\subsection{From Stochastic Processes to Deterministic ODE}
\label{ssec:modeling_diffusion}
The denoising diffusion probabilistic model (DDPM) \cite{DDPM} defines a discrete-time forward Markov process in which Gaussian noise $\mathbf{\epsilon} \sim \mathcal{N}(\mathbf{0}, \mathbf{I})$ 
is incrementally added to data over $T$ steps according to a predefined variance schedule ${\beta}_t$. 
Each intermediate state $\mathbf{x}_t$ can be expressed analytically as a function of the initial data $\mathbf{x}_0$ by:
\begin{equation}
	q(\mathbf{x}_t | \mathbf{x}_0) = \mathcal{N}(\mathbf{x}_t; \sqrt{\bar{\alpha}_t}\mathbf{x}_0, (1-\bar{\alpha}_t)\mathbf{I}),
	\label{eq:ddpm_forward_direct}
\end{equation}
where $\alpha_t = 1 - \beta_t$ and $\bar{\alpha}_t = \prod_{i=1}^t \alpha_i$. 

The corresponding reverse process is modeled by a neural network $\mathbf{\epsilon}_{\mathbf{\theta}}(\mathbf{x}_t, t)$, commonly implemented as an attention-based U-Net, 
which is trained to estimate the noise component $\mathbf{\epsilon}$ from the noisy input $\mathbf{x}_t$.

The inherent randomness of DDPM prevents reliable recovery of embedded secrets and limits its use in steganography. 
To enable reliable secrets extraction, it is necessary to define a deterministic mapping between noise and data. This requirement can be fulfilled by the probability flow ODE \cite{Score_Function}, 
which corresponds to the continuous-time generalization of the diffusion process and preserves the marginal distributions $p_t(\mathbf{x}_t)$ at all times $t$:
\begin{equation}
	\mathrm{d}\mathbf{x}_t = \left[f(\mathbf{x}_t, t) - \frac{1}{2} g^2(t) \nabla_{\mathbf{x}} \log p_t(\mathbf{x}_t) \right] \mathrm{d} t,
	\label{eq:ode_flow}
\end{equation}
where $f(\cdot, \cdot)$ denotes the drift function and $g(\cdot)$ is the diffusion coefficient, both determined by the parameters of the forward diffusion process.
The score function $\nabla_{\mathbf{x}} \log p_t(\mathbf{x}_t)$ can be approximated using the trained denoiser $\mathbf{\epsilon}_{\mathbf{\theta}}(\mathbf{x}_t, t)$ as 
\begin{equation}
	\nabla_{\mathbf{x}} \log p_t(\mathbf{x}_t) \approx -\frac{\mathbf{\epsilon}_{\mathbf{\theta}}(\mathbf{x}_t, t)}{\sigma_t},
	\label{eq:score_function}
\end{equation}
with $\sigma_t = \sqrt{1 - \bar{\alpha}_t}$.

Assuming $f$, $g$, and the score function are smooth, the ODE admits a unique solution for any initial state $\mathbf{x}_T$ at $t=T$.
This defines a trajectory $\mathbf{x}_t$ for $t\in[0,T]$, establishing a bijective mapping between the noise state at $t=T$ and the image state at $t=0$.
Consequently, the final image state $\mathbf{x}_0$ is fully determined by the original noise $\mathbf{x}_T$, and $\mathbf{x}_T$ can be recovered by integrating the ODE backward, 
providing a mathematically grounded link between the noise and image domains.

\subsection{Efficient Integration via Advanced ODE Solvers}
Efficient and accurate integration of the probability flow ODE in Eq.~\cref{eq:ode_flow} is crucial for practical steganography applications. 
The DDIM sampler \cite{DDIM} accelerates sampling via a first-order Euler scheme but suffers from large numerical errors due to its black-box treatment of Eq.~\cref{eq:ode_flow}, compromising reversibility.

In fact, the ODE can be decomposed into a linear term $f(\mathbf{x}_t, t)$ and a nonlinear term $-\tfrac{1}{2} g^2(t) \nabla_{\mathbf{x}} \log p_t(\mathbf{x}_t)$, where the former admits a closed-form solution. 
DPM-Solver samplers \cite{dpm_solver, dpm_solver_++} exploit this structure by analytically integrating the linear component and numerically approximating only the nonlinear term. 
Through a change of variable $t$ to $\lambda$ with $\lambda_t=\log(\alpha_t/\sigma_t)$, it yields an exact integral formulation in an exponential-integrator form suitable for integration by parts:
\begin{align}
\mathbf{x}_t &= e^{\int_s^t f(\tau)\,\mathrm{d}\tau}\,\mathbf{x}_s + \int_s^t \left(e^{\int_\tau^t f(r)\,\mathrm{d}r}\,\frac{g^2(\tau)}{2\sigma_\tau}\,
\mathbf{\epsilon}_\theta(\mathbf{x}_\tau,\tau)\right)\mathrm{d}\tau \notag \\
 	&= \frac{\alpha_t}{\alpha_s}\,\mathbf{x}_s - \alpha_t \int_{\lambda_s}^{\lambda_t} e^{-\lambda}\,
\hat{\mathbf{\epsilon}}_\theta(\hat{\mathbf{x}}_\lambda,\lambda)\,
\mathrm{d}\lambda.
\end{align}

This formulation enables high-order integration through Taylor expansion of the noise prediction $\hat{\mathbf{\epsilon}}_\theta(\hat{\mathbf{x}}_\lambda, \lambda)$ around $\lambda_{t_{i-1}}$, 
combined with integration by parts to derive recursive coefficients. 
Specifically, define the exponentially weighted integral coefficients:
\begin{align}
C_0 &= \int_{\lambda_{t_{i-1}}}^{\lambda_{t_i}} e^{-\lambda}\,\mathrm{d}\lambda = e^{-\lambda_{t_{i-1}}} - e^{-\lambda_{t_i}} \\
C_n &= \int_{\lambda_{t_{i-1}}}^{\lambda_{t_i}} e^{-\lambda}\frac{(\lambda-\lambda_{t_{i-1}})^n}{n!}\,\mathrm{d}\lambda = C_{n-1} - \frac{h_i^n}{n!}e^{-\lambda_{t_i}}, \notag
\end{align}
where $h_i = \lambda_{t_i} - \lambda_{t_{i-1}}$ denotes the step size in the $\lambda$ domain, and a smaller $h_i$ enhance numerical stability and reduce discretization error.

Substituting the Taylor expansion into the exact solution and using these coefficients yields the $k$-th order approximation:
\begin{equation}
	\mathbf{x}_{t_{i-1} \to t_i} \approx \frac{\alpha_{t_i}}{\alpha_{t_{i-1}}} \tilde{\mathbf{x}}_{t_{i-1}} - \alpha_{t_i} \sum_{n=0}^{k-1} \hat{\mathbf{\epsilon}}_\theta^{(n)}(\hat{\mathbf{x}}_{\lambda_{t_{i-1}}}, \lambda_{t_{i-1}}) \cdot C_n,
\end{equation}
where $\hat{\mathbf{\epsilon}}_\theta^{(n)}$ denotes the $n$-th order derivative.

The approximation error decreases from $O(h_i)$ in DDIM to $O(h_i^3)$ per time step in the 3-order DPM-Solver, greatly improving BAR in diffusion-based steganography.

\section{Proposed Method}
\label{sec:proposed_method}
This section presents a steganography framework grounded in the reversibility of ODE-based diffusion, as illustrated in Fig.~\cref{fig:overview}.
Building on an analysis of numerical inversion under prior deviations, we propose the Approximate Gaussian Mapping (AGM) for efficient secret embedding.
Furthermore, a two-stage optimization strategy is introduced to determine the mapping parameters that minimize statistical divergence while ensuring the target recovery bit accuracy rate.

\subsection{Impact of Prior Deviation on Numerical Reversibility}
\label{ssec:deviation_analysis}
Although mapping secrets to the SNP is a central objective in diffusion-based steganography, 
the empirical results reported in prior studies \cite{Position, CLT, DCT_noise, Cross, GSF} indicate that non-SNP inputs can also achieve effective performance.
This motivates a systematic investigation into the effects of intentional deviation. 

We evaluate controlled prior deviations on two open-source diffusion models\footnote{Code is available at \url{https://github.com/LuChengTHU/dpm-solver}}: 
a pixel-space model trained on LSUN Bedroom \cite{Lsun}, and a latent-space text-conditioned model based on Stable Diffusion v1.5 model \cite{SD}, 
using prompts from the ImageInWords (IIW) dataset \cite{ImageInWords} as conditioning input. 
	
The deviation is introduced by increasing the variance of the original noise input. Let $\mathbf{x}_T \sim \mathcal{N}(\mathbf{0},\mathbf{I})$ be the reference noise sample, and define the perturbed input as  
\begin{equation}
    \mathbf{x}_T' = \mathbf{x}_T + \sigma \cdot \mathbf{\epsilon}, \quad \mathbf{\epsilon} \sim \mathcal{N}(\mathbf{0}, \mathbf{I}),
\end{equation}
where $\sigma \in [0, 2]$ controls the perturbation magnitude. 

Each perturbed sample $\mathbf{x}_T'$ is propagated through the diffusion model $\mathcal{DM}$ to produce the corresponding image $\mathbf{x}_0' = \mathcal{DM}(\mathbf{x}_T')$, 
and then inverted back to the noise domain $\hat{\mathbf{x}}_T = \mathcal{DM}^{-1}(\mathbf{x}_0')$. 
The reconstruction error is computed as $\Delta \mathbf{x} = \hat{\mathbf{x}}_T - \mathbf{x}_T'$, from which both the mean and the maximum absolute error are recorded for each perturbation level. 

Image fidelity and steganalysis resistance are evaluated in a later subsection, where candidate distributions are restricted by KL divergence. 
This separation isolates the effect of prior deviation on reversibility at this stage, while the subsequent stage integrates reversibility with perceptual quality and detectability under distributions close to the SNP.

Sampling is consistently executed via the multi-step third-order DPM-Solver++ (50 sampling steps, logSNR schedule) throughout this study, with a guidance scale of 7.5 applied to the SD v1.5 model. 
All experiments are implemented in PyTorch 2.7 on an NVIDIA RTX PRO 6000 GPU.

\begin{figure}[!t]
	\centering
	\subfloat[]{\includegraphics[width=0.45\columnwidth, trim=0 5 0 5, clip]{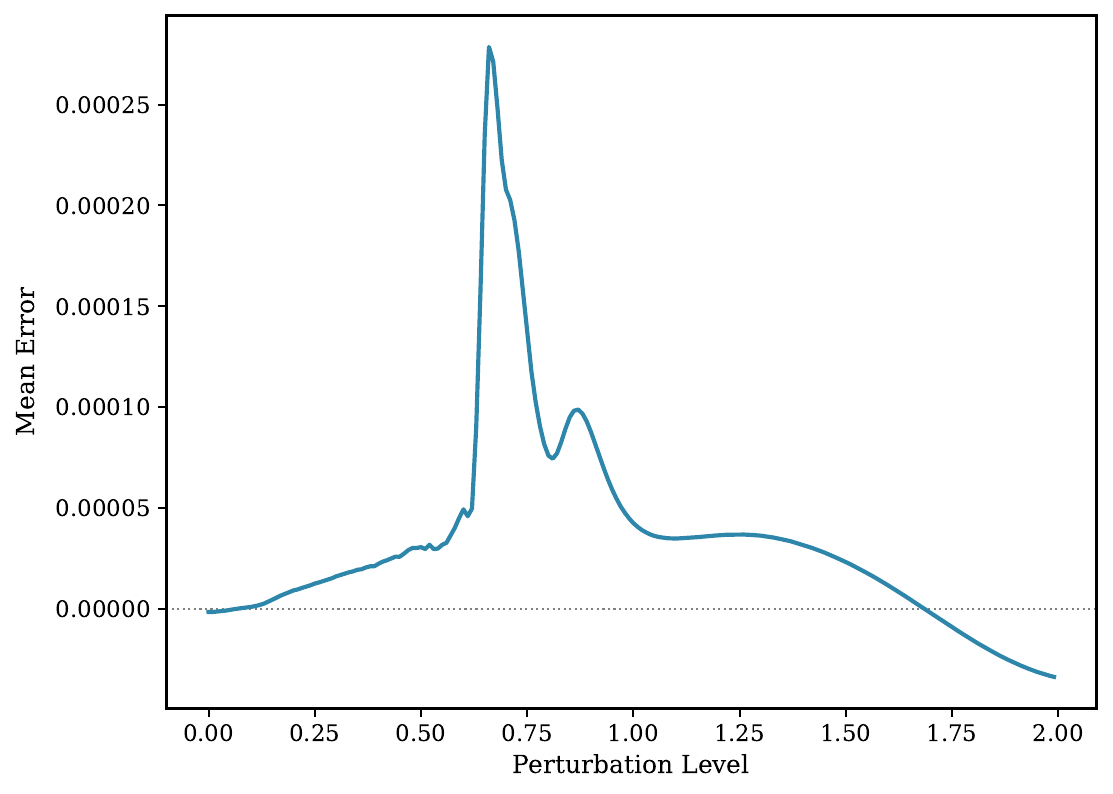}
		\label{fig:pixel_mean_error}}
	\hfil
	\subfloat[]{\includegraphics[width=0.45\columnwidth, trim=0 5 0 5, clip]{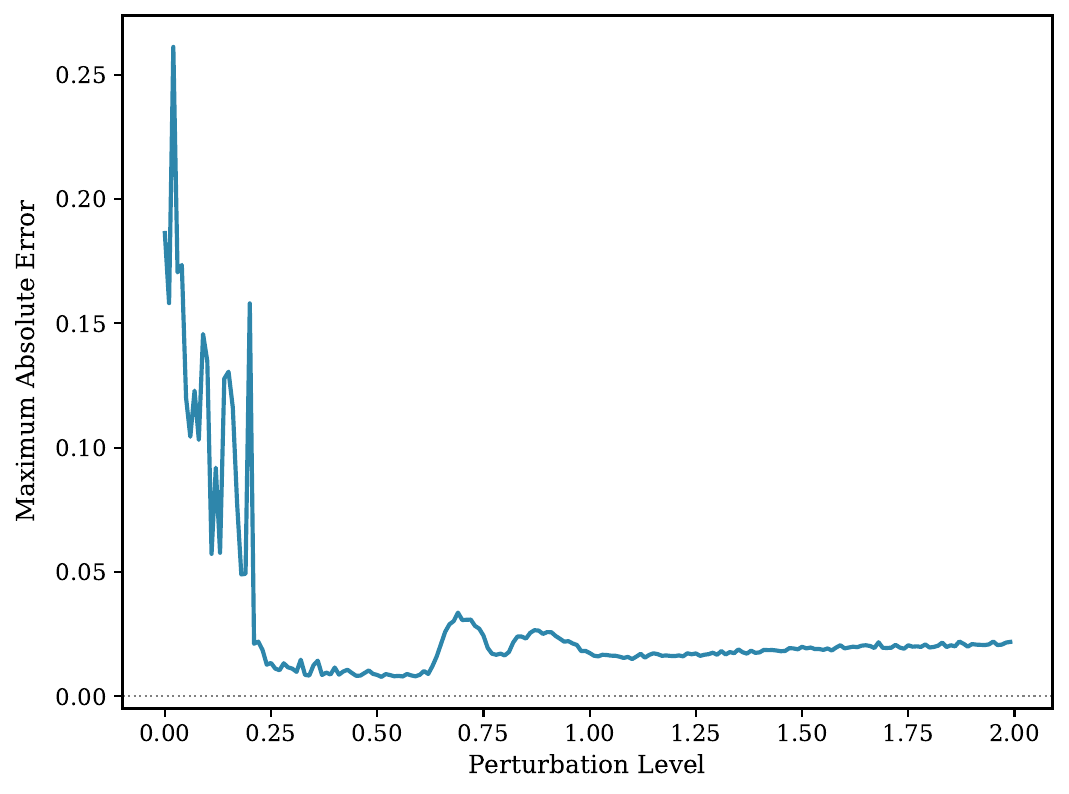}
		\label{fig:pixel_max_error}}
	
	\vspace{-1.0em}
	
	\subfloat[]{\includegraphics[width=0.45\columnwidth, trim=0 5 0 5, clip]{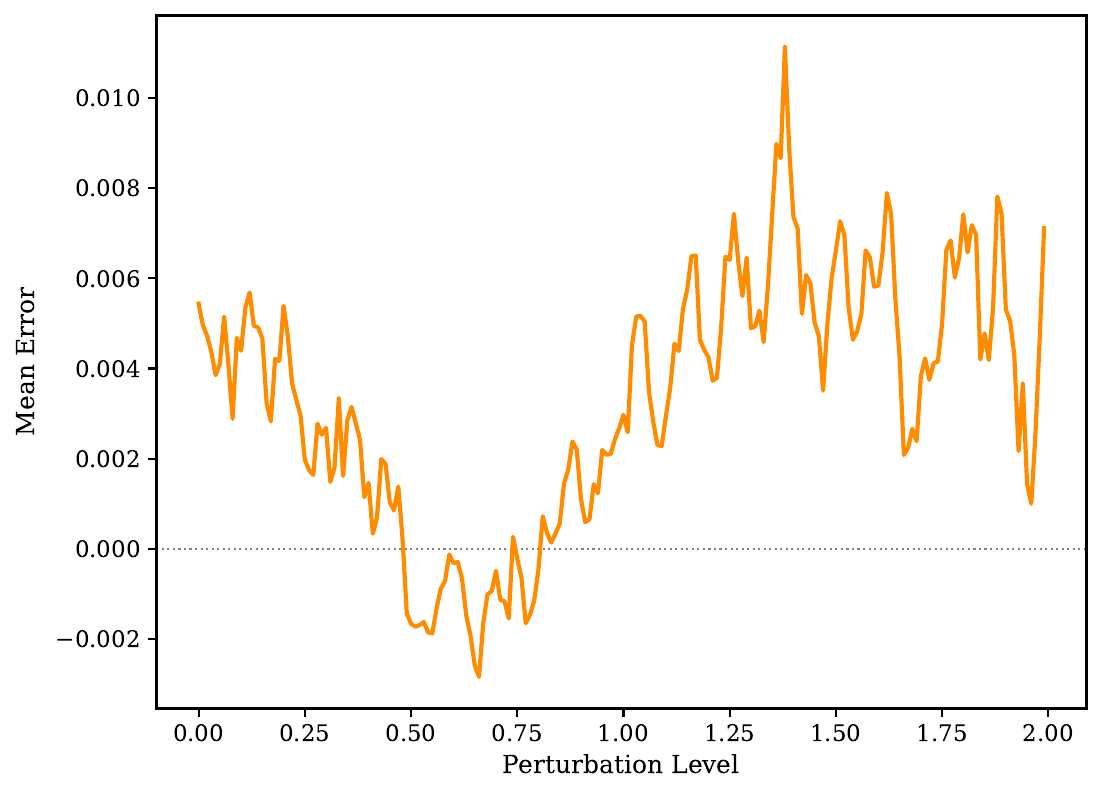}
		\label{fig:sd_mean_error}}
	\hfil
	\subfloat[]{\includegraphics[width=0.45\columnwidth, trim=0 5 0 5, clip]{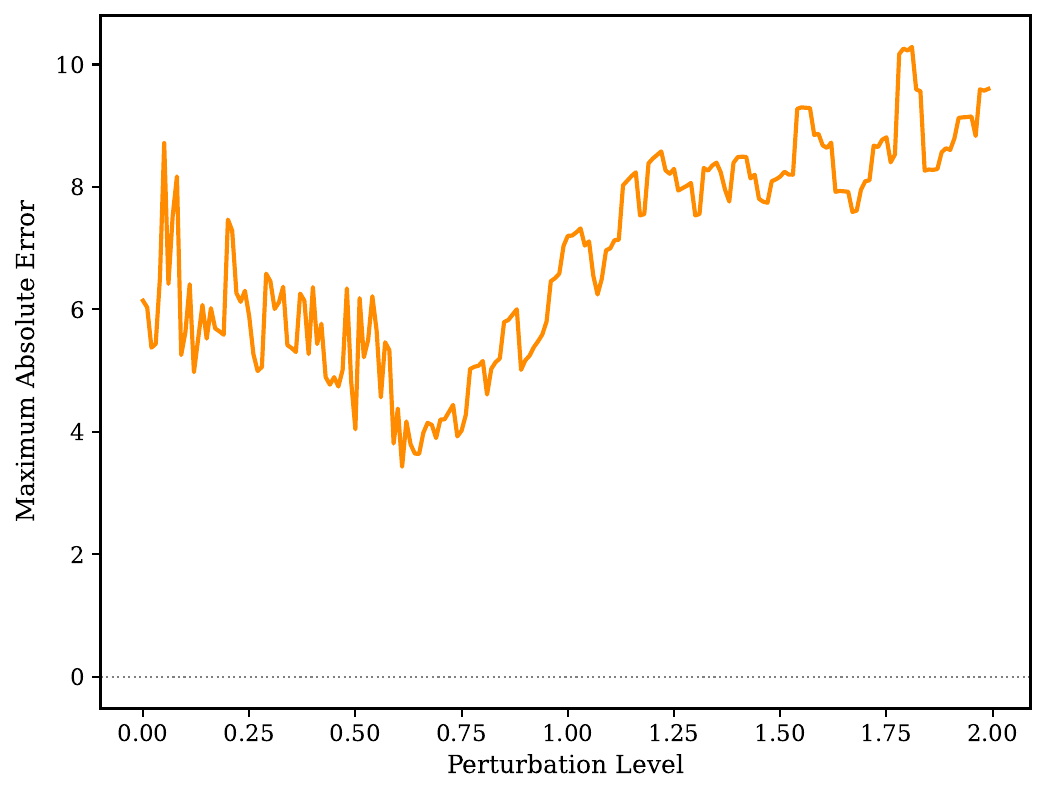}
		\label{fig:sd_max_error}}
	
	\caption{Noise reconstruction error under controlled prior deviation. (a, b) Pixel space and (c, d) latent space, illustrating mean (left) and maximum absolute (right) errors.}
	\label{fig:error_deviation_compare}
\end{figure}

The non-monotonic error profiles in Fig. \ref{fig:error_deviation_compare} show that the SNP does not minimize numerical reconstruction error, 
and that both the mean and the maximum absolute error attain their lower values at a distinct non-zero perturbation level.
This indicates that a controlled departure from the SNP can improve numerical reversibility and motivates the construction of an explicitly regulated approximate Gaussian prior in the next subsection.

\subsection{Approximate Gaussian Mapping}
\label{ssec:AGM}
Building on the findings in \cref{ssec:deviation_analysis} that remove the constraint of strict SNP matching, we propose the Approximate Gaussian Mapping (AGM). 
The AGM is a direct linear transformation designed to evaluate the utility of computationally efficient mappings for steganography.

The secrets embedding procedure partitions a secret bitstream into a sequence of $Q$-bit integers $m$, 
whose discrete uniform distribution is ensured by a cryptographic pre-processing stage such as AES \cite{AES}, which also enhances overall system security.
Each integer is mapped to an initial noise vector component $\mathbf{x}_T$ through a two-step operation. 

First, each Q-bit integer $m$ is mapped to a value $u$ from a zero-mean uniform distribution whose scale is governed by the parameter $S$: 
\begin{equation}
	u = S \cdot \left(\frac{m}{2^Q-1} - 0.5\right).
	\label{eq:agm_gen_u}
\end{equation}

The value $u$ is then additively combined with auxiliary Gaussian noise $n \sim \mathcal{N}(\mathbf{0},\mathbf{I})$, 
which is deterministically sampled with a shared seed $K$, and the composite signal is scaled by the variance factor $\sigma$: 
\begin{equation}
	\mathbf{x}_T = \frac{u+n}{\sigma}.
	\label{eq:agm_gen_xt}
\end{equation}

The complete vector $\mathbf{x}_T$ synthesizes the stego image $\mathbf{x}_{\text{stego}}$ via reverse ODE integration. 
The choice of the scale factor $S$ and the variance factor $\sigma$ critically governs the trade-off between image fidelity, 
BAR, and anti-steganalysis detection, and their optimization is detailed in the subsequent subsection.

An authorized receiver inverts this deterministic process using the shared parameters ($Q, S, \sigma, K$). 
The receiver reconstructs the noise vector $\hat{\mathbf{x}}_T$ from the received stego image $\hat{\mathbf{x}}_{\text{stego}}$ via forward ODE integration and subsequently retrieves the message using the inverse transformation in Eq.~\eqref{eq:agm_recover_m}:
\begin{equation}
	\hat{m} = \left \lfloor \left(\frac{\hat{\mathbf{x}}_{T} \cdot \sigma - n}{S} + 0.5\right) \cdot (2^Q-1)  \right \rceil.
	\label{eq:agm_recover_m}
\end{equation}

The original secret bitstream is then recovered by decrypting the Q-bit sequence formed from the concatenation of the estimated integers $\hat{m}$.

One advantage of the AGM is its favorable time complexity. 
For each $Q$-bit message segment, the online mapping operation has a constant time complexity of O(1). 
This is achieved by pre-computing the optimal parameters $S^*$ and $\sigma^*$ and storing the deterministic signal components $u$ in a lookup table (LUT). 
The online process for each segment $m$ is thus reduced to one LUT, one sampling of the Gaussian noise $n$, and a fixed number of arithmetic operations. 
Consequently, the total time complexity for embedding a secret bitstream of length $L$ is linear, O($L$). 

\subsection{Two-Stage Decoupled Parameter Search}
\label{ssec:optimization}
While the analysis in \cref{ssec:deviation_analysis} indicates that minor deviations from the SNP mitigate numerical inversion errors, 
strict statistical adherence to this prior is widely regarded as a primary condition for security against steganalysis.
To quantify this adherence, we employ the Kullback-Leibler (KL) divergence, a metric intrinsically aligned with the diffusion model's training objective of optimizing the variational lower bound through cumulative KL terms.
Minimizing this divergence ensures that the embedded noise maintains statistical consistency with the distribution learned during training.

To rigorously link the Q-bit per-symbol embedding distortion $D_{KL}^{\text{symbol}}$ to the security of the secret bitstream, we adopt the widely accepted Independent and Identically Distributed (IID) assumption and Gram-Charlier Type A series expansion \cite{Gram–Charlier}. 
Under this assumption, the joint divergence $D_{KL}^{\text{total}}$ scales linearly with the number of Q-bit segments, rendering per-symbol minimization equivalent to global security optimization (refer to Appendix~\ref{app:kl_derivation} for more details).

We therefore formulate the parameter selection as a constrained optimization problem that minimizes the analytical KL divergence approximation subject to a predefined target BAR $\tau$, expressed as
\begin{equation}
	\begin{aligned}
		& \min_{S,\sigma} \quad \mathcal{D}_{\text{KL}}(S, \sigma) \approx \frac{1}{2} \left( \frac{(\kappa_2-1)^2}{2} + \sum_{j=2}^{5} \frac{\kappa_{2j}^2}{(2j)!} \right) \\
		& \text{s.t.} \quad \text{BAR}(S, \sigma) \ge \tau,
	\end{aligned}
	\label{eq:optimization_problem}
\end{equation}
where $\kappa_n$ denotes the $n$-th cumulant of the mapped noise distribution.

Given that the optimization space consists of only two scalar parameters, we adopt a heuristic approach based on the distinct roles of each variable.
The scale factor $S$ determines the inter-symbol distance of the Q-bit message integer, thereby exerting the dominant influence on BAR.
In contrast, $\sigma$ functions as a secondary adjustment that fine-tunes numerical stability.
Leveraging this hierarchical sensitivity, we propose a Two-Stage Decoupled Optimization strategy outlined in \cref{alg:two_stage_optim}.

\begin{algorithm}[!t]
	\caption{Two-Stage Decoupled Parameter Search}
	\label{alg:two_stage_optim}
	\SetKwInOut{Input}{Input}
	\SetKwInOut{Output}{Output}
	\Input{Capacity $Q$, Model $\mathcal{DM}$, Target BAR $\tau$, Scale Range $[S_{min}, S_{max}]$,
	       Phase I Steps $N$, Phase II Steps $K$, Step Size $\Delta S$,
	       Sigma Multipliers $\mathcal{M}_{\sigma} = \{1.0, 1.0 \pm 0.02, 1.0 \pm 0.05\}$}
	\Output{Optimal Parameters $S^*, \sigma^*$}
	\BlankLine
	// Phase I: Binary Search \\
	$S^* \leftarrow \text{None}, \sigma^* \leftarrow \text{None}, \mathcal{D}^* \leftarrow \infty$\\
	$l \leftarrow S_{min}, r \leftarrow S_{max}$\\
	\For{$i \leftarrow 1$ \KwTo $N$}{
		$S \leftarrow (l + r) / 2$, \space $\sigma \leftarrow \sigma_{prior}(S)$, \space $\mathcal{D} \leftarrow \mathcal{D}_{\text{KL}}(S, \sigma)$\\
		\If{$\text{BAR}(S, \sigma) \ge \tau$}{
			$S^* \leftarrow S,\space \sigma^* \leftarrow \sigma, \space \mathcal{D}^* \leftarrow \mathcal{D}$, \space $r \leftarrow S$\\
		}
		\Else{
			$l \leftarrow S$\\
		}
	}
	\BlankLine
	// Phase II: Boundary Refinement \\
	$S_{anchor} \leftarrow S^*$\\
	\For{$k \leftarrow 1$ \KwTo $K$}{
		$S_{cand} \leftarrow S_{anchor} - k \cdot \Delta S$\\
		$\Sigma_{cand} \leftarrow \{ \sigma_{prior}(S_{cand}) \cdot m \mid m \in \mathcal{M}_{\sigma} \}$\\
		\For{$\sigma_{cand} \in \Sigma_{cand}$}{
			$\mathcal{D} \leftarrow \mathcal{D}_{\text{KL}}(S_{cand}, \sigma_{cand})$\\
			\If{$\text{BAR}(S_{cand}, \sigma_{cand}) \ge \tau$ \& $\mathcal{D} < \mathcal{D}^*$}{
				$S^* \leftarrow S_{cand}, \sigma^* \leftarrow \sigma_{cand}, \mathcal{D}^* \leftarrow \mathcal{D}$\\
			}
		}
	}
	\Return{$S^*, \sigma^*$}
\end{algorithm}

Phase I enforces unit variance by deterministically coupling the noise scaling $\sigma$ to the signal scale $S$ via
\begin{equation}
	\sigma_{prior}(S) = \sqrt{1 + \text{Var}(u)} = \sqrt{1 + \frac{S^2(2^Q+1)}{12(2^Q-1)}}.
	\label{eq:sigma_prior}
\end{equation}

In this regime, the BAR exhibits a global monotonic trend with respect to $S$, despite minor stochastic fluctuations.
We employ a binary search algorithm over $[S_{min}, S_{max}]$ to identify the minimum scale factor satisfying the given BAR threshold $\tau$.
The logarithmic convergence rate of this algorithm renders the outcome robust to the initialization width, ensuring rapid contraction to a high-precision solution even with expansive boundary assignments.
This procedure yields an anchor solution $(S^*, \sigma^*)$ that minimizes the KL divergence within this theoretically constrained regime.

Phase II explores the region $S < S^*$ to determine if controlled variance perturbations allow for further divergence minimization.
The algorithm iteratively decreases the candidate $S_{cand}$ and searches for a compensating $\sigma_{cand}$ within a local neighborhood of the prior $\sigma_{prior}(S_{cand})$.
A candidate configuration replaces the current optimal solution if and only if it maintains the required BAR while achieving a strictly lower divergence $\mathcal{D}$.

\begin{table}[htbp]
	\centering
	\caption{The Computed Parameters ($S^*, \sigma^*$) Minimizing Theoretical KL Divergence Subject to Target Bit Accuracy Rate Constraints.}
	\label{tab:s_optimization_data}
	\begin{tabular}{@{}lccccc@{}}
		\toprule
		{}       & $Q$ & $\tau$ & $S^*$ & $\sigma^*$  & $D_{KL}$ \\
		\midrule
		\multirow{4}{*}{Bedroom} 
		& 1   & 99.99\%        & 0.1083    & 1.0015      & 6.0887e-12     \\
		& 2   & 99.00\%        & 0.1204    & 1.0010     & 6.2810e-13     \\
		& 4   & 97.00\%        & 0.3779    & 1.0067      & 9.5596e-10     \\
		& 8   & 95.00\%        & 7.2180    & 2.3186      & 0.1030         \\
		\midrule 
		\multirow{4}{*}{IIW} 
		& 1   & 98.00\%        & 2.1875    & 1.4820      & 0.0390         \\
		& 2   & 90.00\%        & 5.3125    & 2.2181	    & 0.1492         \\
		& 4   & 70.00\%        & 10.0625   & 3.2501      & 0.2703         \\
		& 8   & 60.00\%        & 16.9735   & 5.0196      & 0.4447         \\
		\bottomrule
	\end{tabular}
\end{table}	

\begin{table*}[!t]
	\centering
	\caption{Quantitative comparison of Generated Images FID and Retrieval BAR. Data is presented in the format: Pixel-Space / Latent-Space.}
	\label{tab:cap_fid_comparison}
	
	\scriptsize
	\renewcommand{\arraystretch}{1.5} 
	\setlength{\tabcolsep}{1.5pt}     
	\newcommand{\dt}[2]{#1 \newline #2}
	
	\newcolumntype{C}{>{\centering\arraybackslash}X}
	
	\begin{tabularx}{\textwidth}{@{} l | *{5}{C} | C | C | C C | C C @{}}
		\toprule

		\multirow{2}{*}{\diagbox[width=11em, height=3.5em]{\raisebox{0.5ex}{\textbf{Metric}\,\,}}{\raisebox{0.5ex}{\,\,\textbf{Method}}}} & 
		\multicolumn{5}{c|}{\multirow{2}{*}{\textbf{PROPOSED METHOD}}} & 
		\multicolumn{6}{c}{\textbf{SOTA METHODS}} \\
		\cline{7-12} 
		
		& \multicolumn{5}{c|}{} & 
		IDCT \cite{DCT_noise} & ICDF \cite{Stegaddpm} & \multicolumn{2}{c|}{CLT \cite{CLT}} & \multicolumn{2}{c}{GSF \cite{GSF}} \\	
		\hline
		
		\textbf{Capacity ($Q$)} & 
		0 & 1 & 2 & 4 & 8 & 
		1 & 1 & 1 & 8 & 1 & 8 \\
		\hline
		
		\textbf{FID} $\downarrow$ & 
		\dt{3.13  /  }{24.01} &   
		\dt{3.16  /  }{24.44} &   
		\dt{3.16  /  }{24.82} &   
		\dt{3.21  /  }{25.13} &   
		\dt{3.79  /  }{25.13} &   
		\dt{3.80  /  }{24.04} & 	 
		\dt{3.33  /  }{24.87} & 	 
		\dt{3.17  /  }{24.34} & 	 
		\dt{3.34  /  }{25.18} & 	 
		\dt{3.11  /  }{24.03} & 	 
		\dt{408.25  /  }{62.66} \\	 
		\hline
		
		\textbf{BAR} (\%) $\uparrow$ & 
		\dt{-  /  }{-} &                   
		\dt{99.99  /  }{97.83} & 
		\dt{99.06  /  }{89.84} & 
		\dt{97.04  /  }{69.73} & 
		\dt{95.10  /  }{59.67} & 
		\dt{100.00  /  }{97.30} & 
		\dt{75.00  /  }{73.81} &  
		\dt{100.00  /  }{97.58} & 
		\dt{91.53  /  }{57.74} &  
		\dt{99.80  /  }{86.51} &  
		\dt{94.36  /  }{55.14} \\ 
		
		\bottomrule
	\end{tabularx}
\end{table*}
This decoupled strategy reduces optimization complexity to a logarithmic factor for the anchor search followed by a few number of evaluations for the local refinement.
The found solutions are listed in \cref{tab:s_optimization_data}.
Notably, for embedding capacities $Q \le 4$ under pixel-space architectures, the optimization yields negligible divergence magnitudes, 
ensuring that the secret mapped noise remains statistically indistinguishable from the SNP. 

The disparate target BAR $\tau$ address specific architectural constraints. 
While pixel-space diffusion permits near-perfect reversibility limited only by numerical precision, latent models involve a lossy VAE encoding-decoding cycle. 
This perceptual compression introduces intrinsic information loss, necessitating lower targets for the IIW dataset to accommodate this structural bound.

These parameters are computed offline and remain fixed for all subsequent experiments, imposing no additional overhead during online generation.

\section{Experiments}
\label{sec:experiments}
This section presents the empirical evaluation of the proposed framework.
We first conduct a comparative analysis of various mapping strategies under a unified experimental setting to benchmark fidelity, retrieval bit accuracy rate, and security.
Subsequently, we investigate the architecture-dependent trade-off between security and robustness, providing a mechanistic explanation for the divergent behaviors observed in pixel-space and latent-space models.

\subsection{Experimental Settings}
The evaluation utilizes the diffusion model architectures, datasets, and solver configurations detailed in Section~\ref{ssec:deviation_analysis}.
The effective embedding rate is normalized by architecture.
In pixel-space models, the noise dimensions align with the spatial output, yielding a rate of $Q$ bits per pixel per channel (bpp/ch).
Conversely, the latent-space model operates within a compressed $4 \times 64 \times 64$ representation.
Decoding this tensor to a $3 \times 512 \times 512$ resolution results in a scaled effective rate of $Q \times (4 \cdot 64^2) / (3 \cdot 512^2)$ bpp.

To ensure a rigorous comparison, we do not rely on statistics reported in the original publications of the SOTA methods \cite{Stegaddpm, CLT, DCT_noise, GSF}, which employ disparate model checkpoints and sampling schedulers.
Instead, we re-implemented the core secret-to-noise mapping methods of these baselines and integrated them into the unified diffusion framework described above.
All comparative experiments utilize identical backbones, random seeds, and DPM-Solver++ configurations, thereby isolating the influence of the mapping strategy on performance.

We evaluate the proposed framework based on three criteria: fidelity, recovery bit accuracy rate, and security.
The Fréchet Inception Distance (FID) \cite{FID} quantifies image fidelity by measuring the distributional divergence between generated and reference samples.
The bit accuracy rate (BAR) measures the integrity of the recovered secret bitstream.
Security is assessed via the minimum classification error $P_E$ of steganalysis networks including SRNet \cite{SRNet}, XuNet \cite{XuNet}, YeNet \cite{YeNet}, and SiaStegNet \cite{SiaStegNet}.
These networks are retrained to detect distributional discrepancies between images synthesized from SNP noise and secret-embedded noise.
A $P_E$ value approaching $0.5$ indicates statistical indistinguishability.

\subsection{Analysis of Fidelity and Capacity}
\label{ssec:fidelity_capacity_analysis}
\begin{figure*}[!t]
	\centering
	\setlength{\tabcolsep}{1.5pt} 
	\renewcommand{\arraystretch}{1.1} 	
	\begin{tabular}{cccccc}
		\resizebox{0.162\textwidth}{!}{
			\begin{tikzpicture}[baseline=(img-3.south)]
				\def\xstep{0.05} 
				\def\ystep{-0.12} 
				
				\foreach \i/\filename in {
					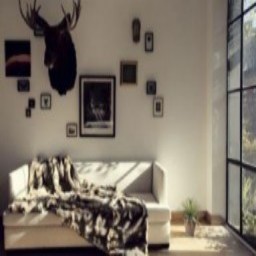,  
					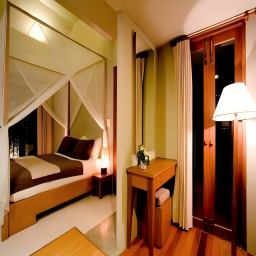,
					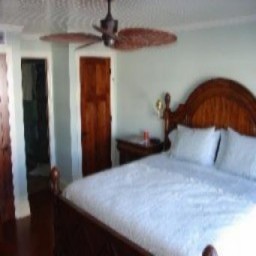,
					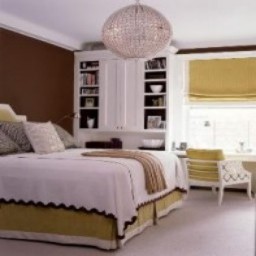     
				} {
					\node[
						name=img-\i,        
						inner sep=0pt,
						anchor=north west,  
						draw=white,         
						line width=2pt,     
						drop shadow={opacity=0.3, shadow xshift=0.5mm, shadow yshift=-0.5mm}
					] at (\i*\xstep, \i*\ystep) { 
						\includegraphics[width=3cm]{\filename}
					};
				}
			\end{tikzpicture}
		} &

		\includegraphics[width=0.162\textwidth]{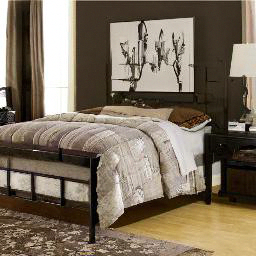} &		
		\includegraphics[width=0.162\textwidth]{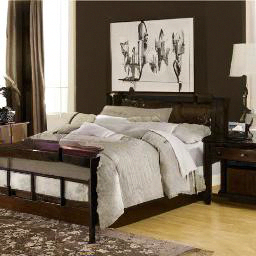} &
		\includegraphics[width=0.162\textwidth]{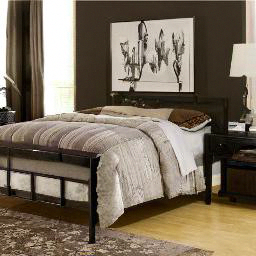} &
		\includegraphics[width=0.162\textwidth]{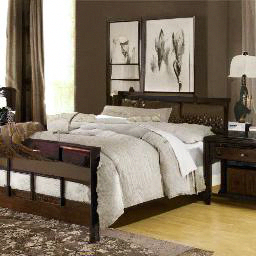} &
		\includegraphics[width=0.162\textwidth]{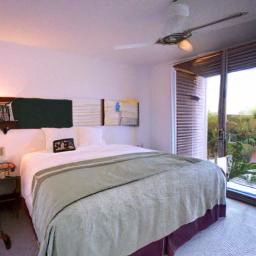} \\		
		\footnotesize Dataset Samples & \footnotesize Non-Stego & \footnotesize AGM ($Q=1$) & \footnotesize AGM ($Q=2$) & \footnotesize AGM ($Q=4$) & \footnotesize AGM ($Q=8$) \\
		
		\includegraphics[width=0.162\textwidth]{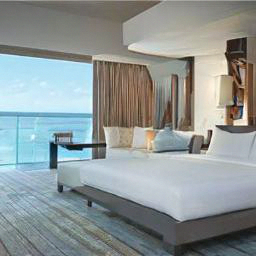} &
		\includegraphics[width=0.162\textwidth]{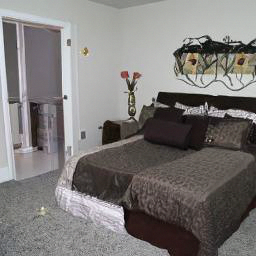} &
		\includegraphics[width=0.162\textwidth]{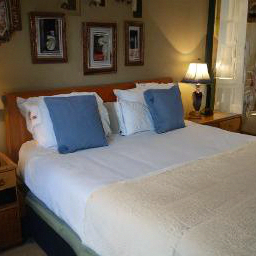} &
		\includegraphics[width=0.162\textwidth]{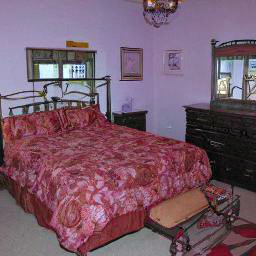} &
		\includegraphics[width=0.162\textwidth]{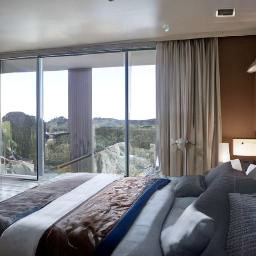} &
		\includegraphics[width=0.162\textwidth]{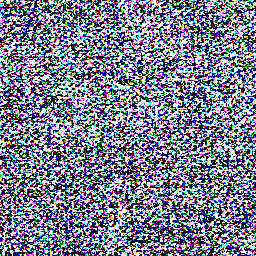} \\
		\footnotesize IDCT ($Q=1$) & \footnotesize ICDF ($Q=1$) & \footnotesize CLT ($Q=1$) & \footnotesize CLT ($Q=8$) & \footnotesize GSF ($Q=1$) & \footnotesize GSF ($Q=8$) \\
	\end{tabular}
	
	\caption{Visual fidelity comparison on LSUN Bedroom (Pixel-Space). 
	\textbf{Top:} Exemplary samples from LSUN Bedroom dataset (stacked), non-stego image generated from the unperturbed SNP noise, and proposed AGM at varying capacities. \textbf{Bottom:} Comparative baseline methods.}
	\label{fig:bedroom_visual}
\end{figure*}

\begin{figure*}[!t]
	\centering
	\setlength{\tabcolsep}{1.5pt} 
	\renewcommand{\arraystretch}{1.1} 	
	\begin{tabular}{cccccc}
		\resizebox{0.162\textwidth}{!}{
			\begin{tikzpicture}[baseline=(img-3.south)]
				\def\xstep{0.05} 
				\def\ystep{-0.12} 
				
				\foreach \i/\filename in {
					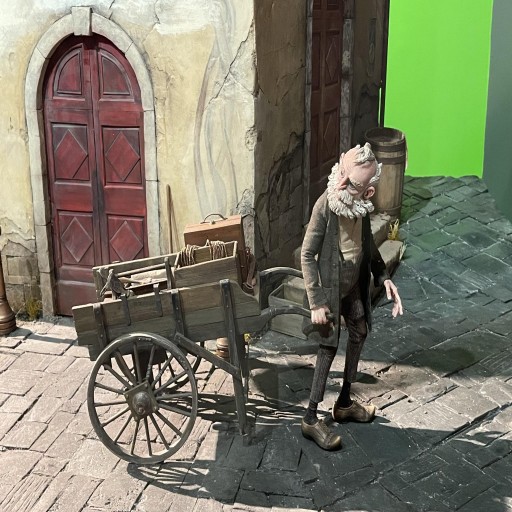,  
					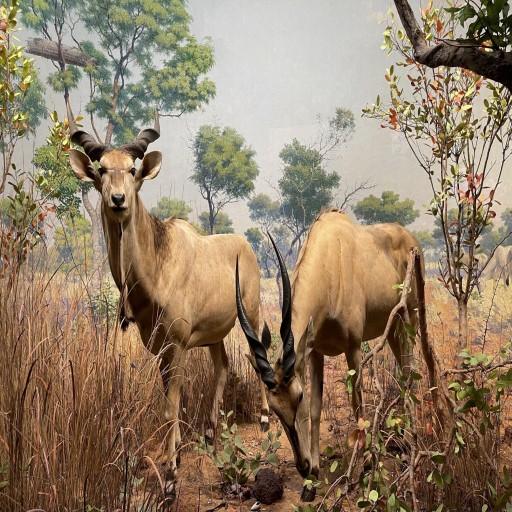,
					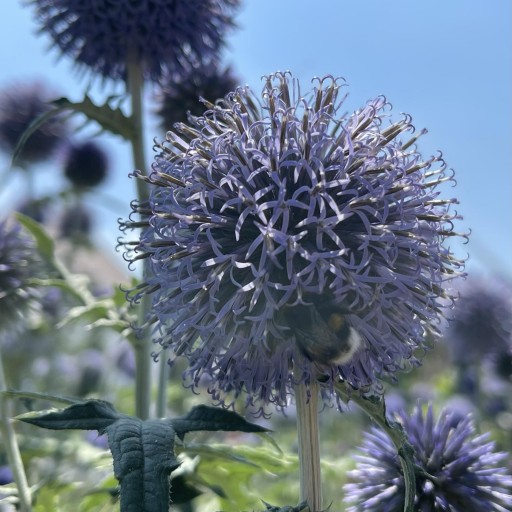,
					     
				} {
					\node[
						name=img-\i,        
						inner sep=0pt,
						anchor=north west,  
						draw=white,         
						line width=2pt,     
						drop shadow={opacity=0.3, shadow xshift=0.5mm, shadow yshift=-0.5mm}
					] at (\i*\xstep, \i*\ystep) { 
						\includegraphics[width=3cm]{\filename}
					};
				}
			\end{tikzpicture}
		} &	
		\includegraphics[width=0.162\textwidth]{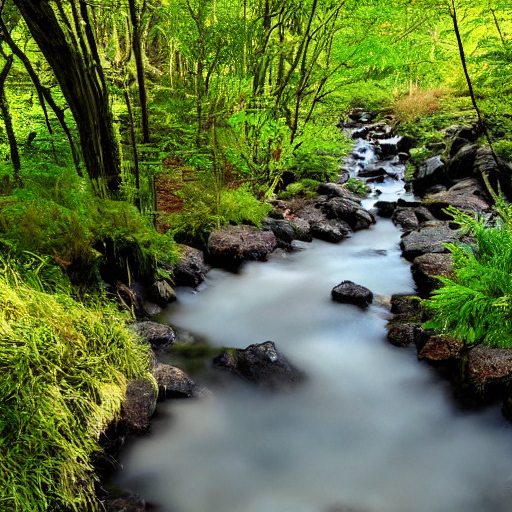} &		
		\includegraphics[width=0.162\textwidth]{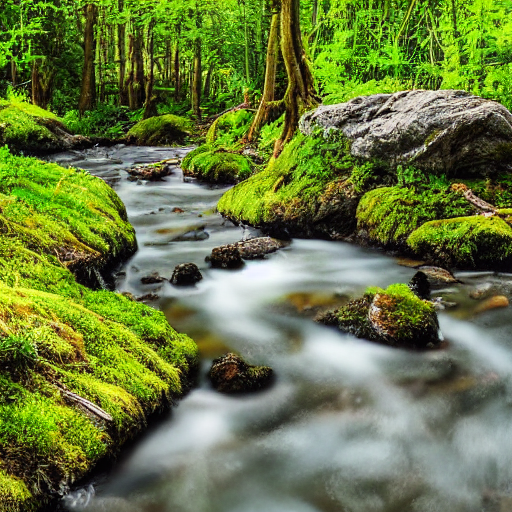} &
		\includegraphics[width=0.162\textwidth]{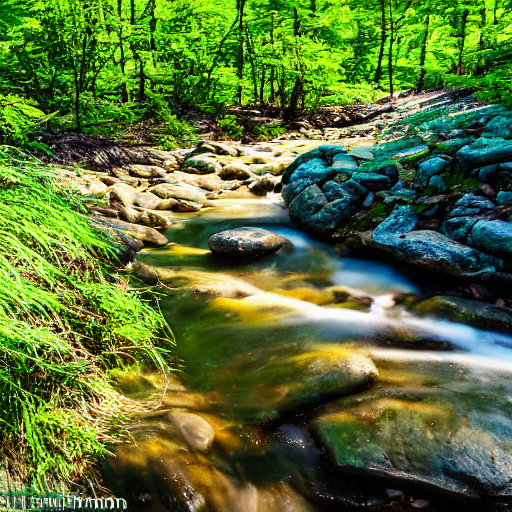} &
		\includegraphics[width=0.162\textwidth]{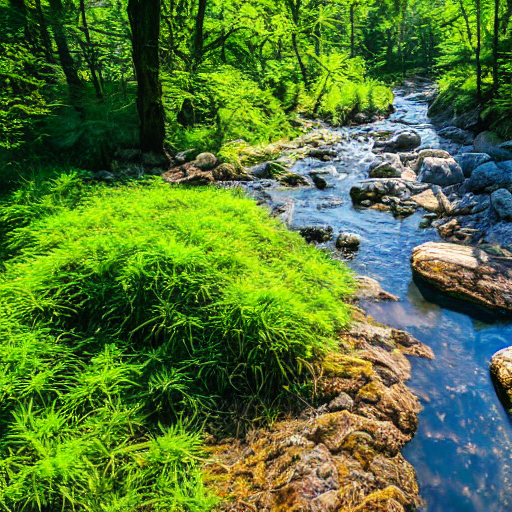} &
		\includegraphics[width=0.162\textwidth]{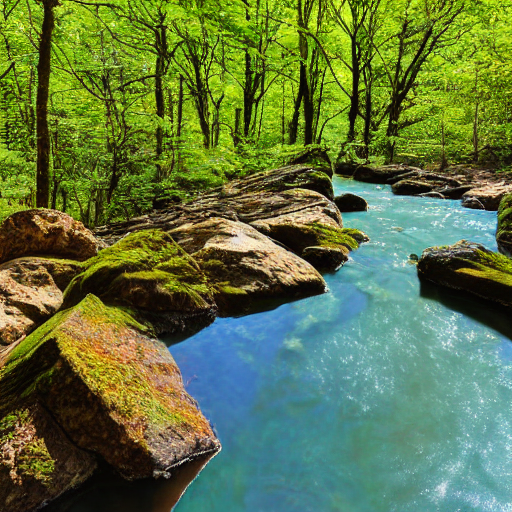} \\		
		\footnotesize Dataset Samples & \footnotesize Non-Stego & \footnotesize AGM ($Q=1$) & \footnotesize AGM ($Q=2$) & \footnotesize AGM ($Q=4$) & \footnotesize AGM ($Q=8$) \\
		
		\includegraphics[width=0.162\textwidth]{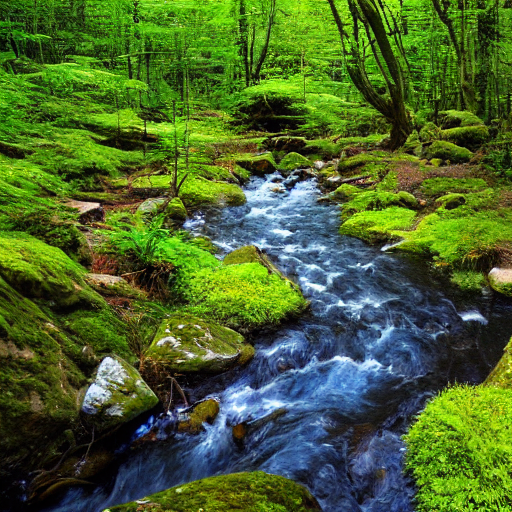} &
		\includegraphics[width=0.162\textwidth]{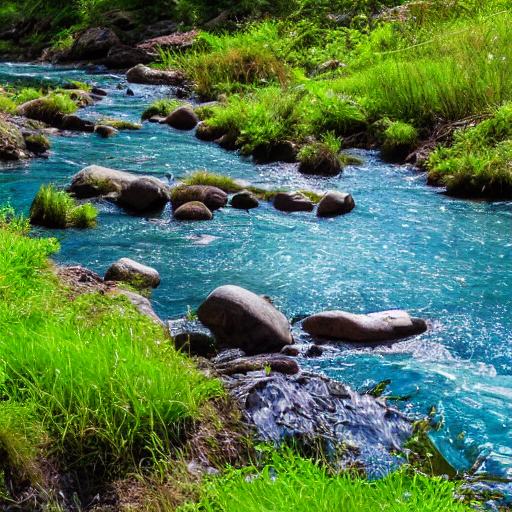} &
		\includegraphics[width=0.162\textwidth]{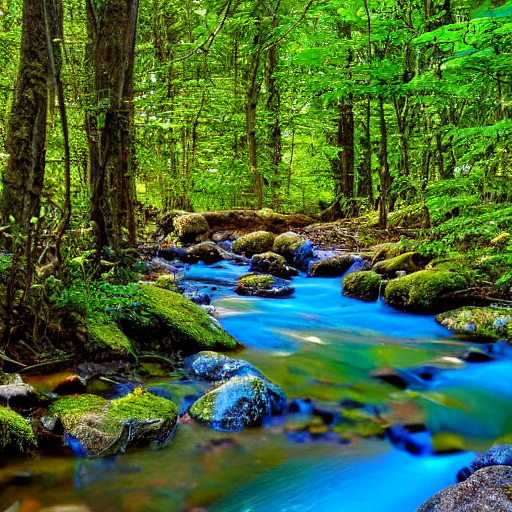} &
		\includegraphics[width=0.162\textwidth]{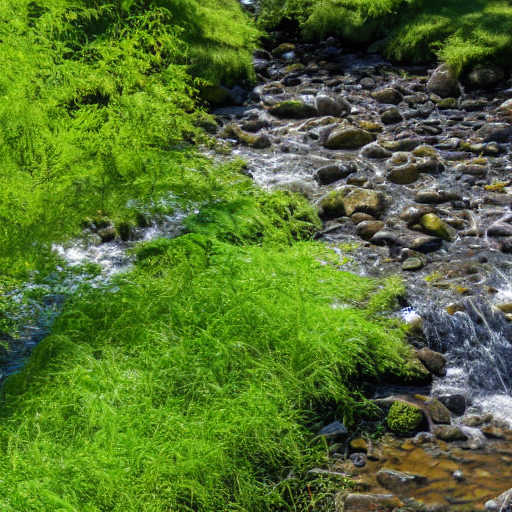} &
		\includegraphics[width=0.162\textwidth]{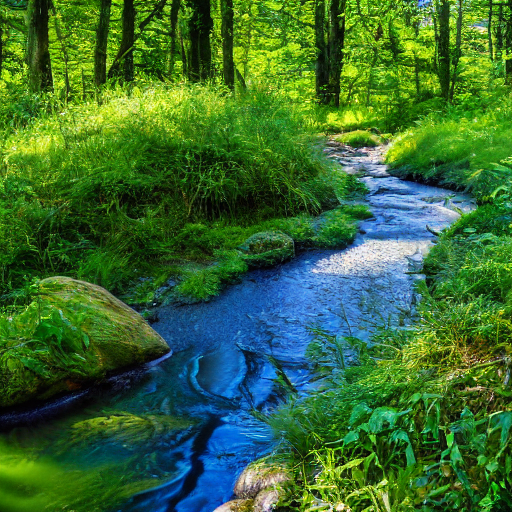} &
		\includegraphics[width=0.162\textwidth]{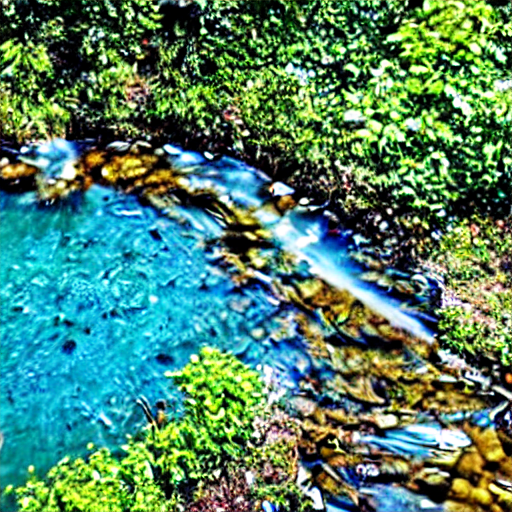} \\
		\footnotesize IDCT ($Q=1$) & \footnotesize ICDF ($Q=1$) & \footnotesize CLT ($Q=1$) & \footnotesize CLT ($Q=8$) & \footnotesize GSF ($Q=1$) & \footnotesize GSF ($Q=8$) \\
	\end{tabular}
	
	\caption{Visual fidelity comparison using the SD v1.5 model (Latent-Space) conditioned on an IIW prompt describing a stream meandering through a verdant forest. 
	\textbf{Top:} Exemplary samples from IIW dataset (stacked), non-stego image generated from the unperturbed SNP noise, and proposed AGM at varying capacities. \textbf{Bottom:} Comparative baseline methods.}
	\label{fig:IIW_visual}
\end{figure*}

Qualitative assessment in Fig. \cref{fig:bedroom_visual} and Fig. \cref{fig:IIW_visual} indicates that AGM achieves high visual quality.
In the pixel-space domain, settings with $Q \le 4$ yield outputs structurally identical to the unperturbed baseline, consistent with the negligible theoretical divergence, 
while the $Q=8$ setting induces a mode shift consistent with trajectory bifurcation. 
Conversely, GSF \cite{GSF} reveals distinct architecture-specific sensitivity by collapsing into noise in pixel-space models yet retaining distorted semantic content in latent-space architectures.


Quantitative analysis is presented in \cref{tab:cap_fid_comparison}, where the $Q=0$ configuration serves as the baseline, representing non-stego generation derived from the SNP noise.

The empirical consistency between the target constraints in \cref{tab:s_optimization_data} and the retrieval results in \cref{tab:cap_fid_comparison} corroborates the efficacy of the proposed optimization strategy.
Specifically, pixel-space models achieve strict adherence to the pre-set BAR targets, while latent-space models converge to the optimal parameters within the boundaries set by target BAR.
Crucially, the analysis reveals a non-linear scaling between statistical KL divergence and perceptual FID degradation.
As evidenced by the pixel-space data, while the KL divergence at $Q=8$ increases by orders of magnitude, the corresponding FID degradation is limited to $19.94\%$.

Regarding visual fidelity, GSF \cite{GSF} at $Q=1$ yields the lowest FID scores, strictly adhering to the performance of the SNP
Conversely, StegaDDPM \cite{Stegaddpm}, despite theoretically mapping to the same prior via the inverse cumulative distribution function (ICDF), exhibits inferior fidelity.
This degradation arises from the numerical instability of the ICDF at distribution tails, where inputs approaching zero or one map to extreme values.
These numerical outliers constitute out-of-distribution inputs for the pre-trained denoising network, causing generation artifacts.
AGM achieves fidelity comparable to orthogonal baselines by balancing statistical alignment with numerical robustness.

In the regime of low numerical deviation (specifically pixel-space at $Q=1$), orthogonal transformation-based methods (IDCT \cite{DCT_noise}, CLT \cite{CLT}) achieve perfect retrieval.
This resilience is intrinsic to the isometric property of orthogonal matrices, which preserves the Euclidean distance of embedded signals, effectively mitigating the minor discretization errors introduced by the ODE solver.
However, this fixed error tolerance becomes insufficient when the system encounters elevated numerical deviations, such as those inherent to the latent-space architecture.
The primary limitation of these SOTA baselines is their fixed transformations that cannot scale the signal constellation to overcome increased interference.
In contrast, AGM overcomes this limitation by leveraging the acceptable BAR negotiation process defined in \cref{tab:s_optimization_data}.
By adaptively amplifying the signal scale to establish a sufficient safety margin against larger numerical deviations, AGM secures superior BAR compared to rigid mapping methods.
This strategy is theoretically grounded in the robust mathematical structure of the diffusion process; indeed, 
the observation that GSF ($Q=8$) maintains a BAR comparable to AGM and CLT despite catastrophic fidelity degradation verifies the intrinsic invertibility of the probability flow ODE, 
demonstrating that the deterministic mapping remains mathematically valid even when the signal distribution deviates significantly from the training prior.

\subsection{Analysis of Security and Robustness}
\label{ssec:dichotomy_analysis}
\paragraph{Security Analysis}
The experimental dataset comprises 20,000 cover images synthesized from SNP noise. Correspondingly, an independent set of 20,000 stego images was generated for every distinct combination of method and embedding capacity $Q$. 
For each experimental configuration, the resulting 40,000 images are pooled and randomly shuffled before being split into training, validation, and testing sets following a $0.70:0.15:0.15$ ratio for steganalysis evaluation.
\begin{table*}[!t]
	\centering
	\caption{Steganalysis performance ($P_E$) comparison. Data is presented in the format: Pixel-Space / Latent-Space. Values closer to \textbf{0.50} indicate higher security.}
	\label{tab:steganalysis}
	
	\scriptsize
	\renewcommand{\arraystretch}{1} 
	\setlength{\tabcolsep}{2pt}       
	
	\newcommand{\dt}[2]{\begin{tabular}{@{}c@{}}#1\\[-0.2em]#2\end{tabular}}
	
	\newcolumntype{C}{>{\centering\arraybackslash}X}
	
	\begin{tabularx}{\textwidth}{@{} l | *{4}{C} | C | C | C C | C @{}}
		\toprule
		
		\multirow{2.5}{*}{\diagbox[width=8em, height=4em]{\raisebox{0.5ex}{\textbf{Detector}\,\,}}{\raisebox{0.5ex}{\,\,\textbf{Method}}}} & 
		\multicolumn{4}{c|}{\textbf{PROPOSED METHOD}} & 
		\multicolumn{5}{c}{\textbf{SOTA METHODS}} \\
		\cmidrule(l){2-10}

		& \multicolumn{4}{c|}{AGM} & IDCT \cite{DCT_noise} & ICDF \cite{Stegaddpm} & \multicolumn{2}{c|}{CLT \cite{CLT}} & GSF \cite{GSF}\\
		& Q = 1 & Q = 2 & Q = 4 & Q = 8  & Q = 1 & Q = 1 & Q = 1 & Q = 8 & Q = 1 \\
		\midrule
		
		SRNet \cite{SRNet} & 
		\dt{0.4873 / }{0.4713} & \dt{0.4880 / }{0.4566} & \dt{0.4867 / }{0.4540} & \dt{0.3227 / }{0.4476} & \dt{0.3102 / }{0.4644} & \dt{0.4766 / }{0.4733} & \dt{0.4753 / }{0.4811} & \dt{0.4667 / }{0.4803} & \dt{0.4733 / }{0.4752} \\
		\midrule
		
		XuNet \cite{XuNet} & 
		\dt{0.4940 / }{0.4606} & \dt{0.4943 / }{0.4583} & \dt{0.4850 / }{0.4410} & \dt{0.2930 / }{0.4362} & \dt{0.1686 / }{0.4815} & \dt{0.4907 / }{0.4881} & \dt{0.4953 / }{0.4971} & \dt{0.4807 / }{0.4856} & \dt{0.4906 / }{0.4930} \\
		\midrule
		
		YeNet \cite{YeNet} & 
		\dt{0.4958 / }{0.4820} & \dt{0.4980 / }{0.4793} & \dt{0.4887 / }{0.4614} & \dt{0.2873 / }{0.4527} & \dt{0.2635 / }{0.4909} & \dt{0.4966 / }{0.4900} & \dt{0.4896 / }{0.4883} & \dt{0.4805 / }{0.4870} & \dt{0.4938 / }{0.4852} \\
		\midrule
		
		SiaStegNet \cite{SiaStegNet} & 
		\dt{0.4773 / }{0.4660} & \dt{0.4840 / }{0.4413} & \dt{0.4693 / }{0.4367} & \dt{0.2107 / }{0.4220} & \dt{0.1703 / }{0.4724} & \dt{0.4719 / }{0.4693} & \dt{0.4793 / }{0.4850} & \dt{0.4773 / }{0.4792} & \dt{0.4911 / }{0.4807} \\
		
		\bottomrule
	\end{tabularx}
\end{table*}

The quantitative data in \cref{tab:steganalysis} highlights a distinct dissociation between visual fidelity and security. 
This phenomenon is particularly evident when analyzing AGM ($Q=8$) and IDCT ($Q=1$) in the pixel domain. 
Although these two configurations exhibit the inferior FID values among the evaluated schemes, the quantitative degradation remains minor compared to the baselines while the resistance to steganalysis collapses. 
The vulnerability arises from different mechanisms in each case. 
Regarding IDCT ($Q=1$), the observed high detectability aligns with the findings in the original study \cite{DCT_noise}, where the authors ascribed the performance deficit to spectral artifacts introduced by the IDCT operation. 
In parallel, the detectability of AGM ($Q=8$) correlates with an enlarged KL divergence that allows the analyzer to distinguish the generated images from the target noise distribution.

Further inspection of the statistical metrics reveals inconsistencies that require deeper scrutiny. 
While the security of the proposed AGM in the pixel space generally improves as the distribution approaches the SNP, GSF ($Q=1$) and ICDF ($Q=1$) theoretically satisfy the SNP distribution yet fail to consistently achieve the best $P_E$ values. 
Moreover, the AGM configurations in the latent space at capacities from $Q=2$ to $Q=8$ exhibit divergence values significantly larger than the pixel-space AGM ($Q=8$) instance but maintain superior security. 
This inversion implies that the interaction between generative noise distributions and the decision boundaries of analyzer involves complex factors beyond theoretical alignment, 
and the specific causes underlying this discrepancy warrant further investigation.

Comparative analysis with SOTA methods provides context for the applicability of the proposed approach. 
It is observed that CLT ($Q=8$) maintains strong security performance in both pixel and latent spaces at high embedding rates. 
While the proposed AGM exhibits sensitivity to high payloads, it achieves near-optimal security at low and medium capacities. 
This performance profile effectively validates the central premise of this work that approximating the SNP constitutes a sufficient condition for effective steganography under reasonable capacity constraints, 
even if high-rate implementations require further optimization to match the security of methods like CLT.

\paragraph{Robustness Analysis}
Robustness against lossy transmission is also a critical metric for steganography utility. 
This subsection evaluates BAR under four distinct attack modalities: additive white Gaussian noise ($AWGN$), JPEG compression, Salt-and-Pepper noise ($S\&P$), and  Gaussian blurring ($G-Blur$), as detailed in \cref{tab:robustness}. 

\begin{table*}[!t]
	\centering
	\caption{Robustness comparison under various attacks. Data is presented in the format: Pixel-Space / Latent-Space. All values represent BAR in percentage.}
	\label{tab:robustness}
	
	\scriptsize
	\setlength{\tabcolsep}{2pt}
	\renewcommand{\arraystretch}{1.3}
	
	\newcolumntype{C}{>{\centering\arraybackslash}X}
	
	\begin{tabularx}{\textwidth}{@{} l | *{4}{C} | C | C | C C | C C @{}}
		\toprule
		
		\multirow{2}{*}{\diagbox[width=6em, height=4.5em]{\raisebox{0.5ex}{\textbf{Attack}\,\,}}{\raisebox{0.5ex}{\,\,\textbf{Method}}}} & 
		\multicolumn{4}{c|}{\textbf{PROPOSED}} & 
		\multicolumn{6}{c}{\textbf{SOTA METHODS}} \\
		\cline{2-11}
		
		& \multicolumn{4}{c|}{AGM} & IDCT \cite{DCT_noise} & ICDF \cite{Stegaddpm} & \multicolumn{2}{c|}{CLT \cite{CLT}} & \multicolumn{2}{c}{GSF \cite{GSF}} \\
		& Q=1 & Q=2 & Q=4 & Q=8 & Q=1 & Q=1 & Q=1 & Q=8 & Q=1 & Q=8 \\
		\hline
		
		\multicolumn{11}{l}{\textit{AWGN}} \\
		$0.001$ & 
		 74.15 / 93.10 		& 61.76 / 82.97 	& 61.77 / 66.85 	& 65.09 / 58.66 	& 100.00 / 97.48 		& 75.00 / 73.86 	& 100.00 / 97.42 		& 67.40 / 57.95 		& 96.74 / 86.79 	& 90.57 / 55.16 \\
		
		$0.01$ & 
		52.56 / 91.67 		& 51.31 / 79.80 	& 51.91 / 65.69 	& 53.99 / 58.39 	& 85.89 / 95.49 		& 69.13 / 73.20 	& 85.07 / 95.44 		& 54.83 / 57.22 		& 76.01 / 84.95 	& 74.07 / 55.09 \\

		$0.1$ & 
		50.33 / 76.33 		& 50.17 / 68.02 	& 50.25 / 60.14 	& 50.69 / 55.02 	& 56.67 / 79.68 		& 54.27 / 66.73 	& 56.71 / 79.16			& 50.97 / 54.14 		& 55.27 / 72.81 	& 59.06 / 54.53 \\
		\hline
		
		\multicolumn{11}{l}{\textit{JPEG}} \\
		$90$ & 
		50.24 / 88.67 		& 50.12 / 78.21 	& 50.18 / 64.79 	& 50.51 / 57.52 	& 55.21 / 93.80 		& 53.06 / 72.45		& 55.03 / 94.15 		& 50.72 / 56.59 	& 54.15 / 83.49 	& 52.68 / 54.88 \\
		
		$70$ & 
		50.20 / 82.29 		& 50.10 / 73.22 	& 50.15 / 61.78 	& 50.41 / 56.26 	& 54.27 / 87.69 		& 52.51 / 69.64 	& 54.12 / 87.46 		& 50.60 / 55.09 	& 53.36 / 78.64 	& 52.35 / 54.80 \\
		
		$50$ & 	
		50.16 / 79.21 		& 50.08 / 70.69 	& 50.13 / 60.59 	& 50.35 / 55.45 	& 53.69 / 84.43 		& 52.10 / 68.63 	& 53.38 / 82.19 		& 50.49 / 54.47 	& 52.98 / 75.29 	& 51.78 / 54.62 \\
		\hline
		
		\multicolumn{11}{l}{\textit{S\&P}} \\
		$1\%$ & 
		50.70 / 69.94 		& 50.36 / 62.90 	& 50.53 / 57.12 	& 51.33 / 53.59 	& 59.21 / 72.96 		& 54.92 / 63.46 	& 58.99 / 69.24 		& 51.34 / 52.84 	& 67.87 / 67.15 	& 67.96 / 54.04 \\
		
		$3\%$ & 
		50.39 / 65.90 		& 50.20 / 59.88 	& 50.29 / 55.53 	& 50.77 / 53.44 	& 55.39 / 66.97 		& 52.40 / 60.91 	& 55.34 / 67.32 		& 50.77 / 52.52 	& 62.37 / 64.39 	& 61.78 / 53.68 \\
		
		$5\%$ & 
		50.30 / 63.79 		& 50.15 / 59.34 	& 50.23 / 55.00 	& 50.61 / 52.93 	& 54.27 / 66.57 		& 51.81 / 60.34 	& 54.20 / 65.72 		& 50.61 / 52.40 	& 60.32 / 63.11 	& 59.39 / 53.56 \\
		\hline
		
		\multicolumn{11}{l}{\textit{G-Blur}} \\
		$3\times3$ & 
		50.94 / 81.36 		& 50.48 / 78.63 	& 50.72 / 64.28 	& 51.77 / 57.51 	& 67.92 / 94.32 		& 60.03 / 72.52 	& 66.39 / 94.11 		& 52.22 / 56.70 	& 64.66 / 83.07 	& 51.99 / 54.95 \\
		
		$5\times5$ & 
		50.29 / 73.15 		& 50.15 / 71.94 	& 50.23 / 60.78 	& 50.61 / 56.01 	& 56.24 / 89.34			& 53.70 / 70.42		& 55.81 / 87.94 		& 50.81 / 55.53 	& 55.75 / 79.38 	& 50.62 / 54.59 \\
		
		$7\times7$ & 
		50.13 / 67.22 		& 50.06 / 68.12 	& 50.10 / 59.27 	& 50.28 / 55.42 	& 52.70 / 83.57			& 51.69 / 67.74 	& 52.63 / 82.37 		& 50.36 / 54.42 	& 52.79 / 74.86 	& 50.34 / 54.12 \\	
		
		\bottomrule
	\end{tabularx}
\end{table*}

A distinct disparity exists between the stability against initial noise perturbation observed in Fig. \cref{fig:error_deviation_compare} and the fragility under image-domain distortion. 
While the image generative process tolerates minor deviations in the noise input, perturbations applied to the generated image cause the reverse Probability Flow ODE to diverge significantly from the expected trajectory. 
This sensitivity precipitates a catastrophic failure in the pixel domain, where the BAR converges to the random-guessing baseline of 50\%. 
Conversely, the latent-space implementation demonstrates superior resilience under identical attack configurations. 
This stability suggests that the VAE encoder functions effectively as a denoising filter during the projection of distorted imagery back into the latent representation.

Comparative analysis with SOTA methods reveals structural distinctions in robustness profiles. The experimental data for IDCT ($Q=1$) and CLT ($Q=1$) exhibit comparable trends. 
This similarity likely derives from their shared mathematical reliance on orthogonal transformations, whether in the frequency or spatial domain, which inherently preserve signal energy against specific linear distortions. 
In contrast, the proposed AGM prioritizes steganography security over robust transmission. The parameter selection strategy for $(S, \sigma)$ specifically targets the minimization of KL divergence to satisfy lossless transmission constraints. 
This optimization inherently discards the redundancy required to withstand channel distortion, resulting in lower BAR scores compared to the other methods.

Consequently, the operational scope of the proposed method aligns with the insights presented in \cite{DCT_noise}. 
The sensitivity of the inverse ODE trajectory to external perturbations indicates that the pixel-space implementation is best positioned for lossless transmission channels, 
whereas the resilience demonstrated by the latent-space approach renders it an essential strategy for open-channel environments where robustness against signal distortion is prioritized.

\subsection{Impact of Semantic Conditioning and VAE on Performance}
\label{ssec:mechanism_analysis}
\paragraph{Spectral Analysis and the Hypothesis of Semantic Masking}
The attribution of IDCT vulnerability strictly to frequency-domain artifacts is challenged by its divergent behavior across spatial domains. 
Given that IDCT inherently modifies spectral coefficients, a critical question arises: why does the latent-space implementation evade the detection patterns observed in the pixel space? 
Furthermore, the proposed AGM ($Q=8$) manifests a parallel performance discrepancy between spatial domains despite the complete absence of frequency-based transformations. 
Does this consistent trend imply that the mechanism governing the enhanced security functions independently of the specific secrets to noise mapping strategy?
This observation narrows the inquiry to the generative architecture itself. Does the semantic conditioning provided by the CLIP text encoder \cite{CLIP} constitute the primary factor for undetectability, 
or does the VAE decoder exert a dominant structural regularization that suppresses statistical anomalies? 
\begin{figure}[!t]
    \centering
    \includegraphics[width=\linewidth]{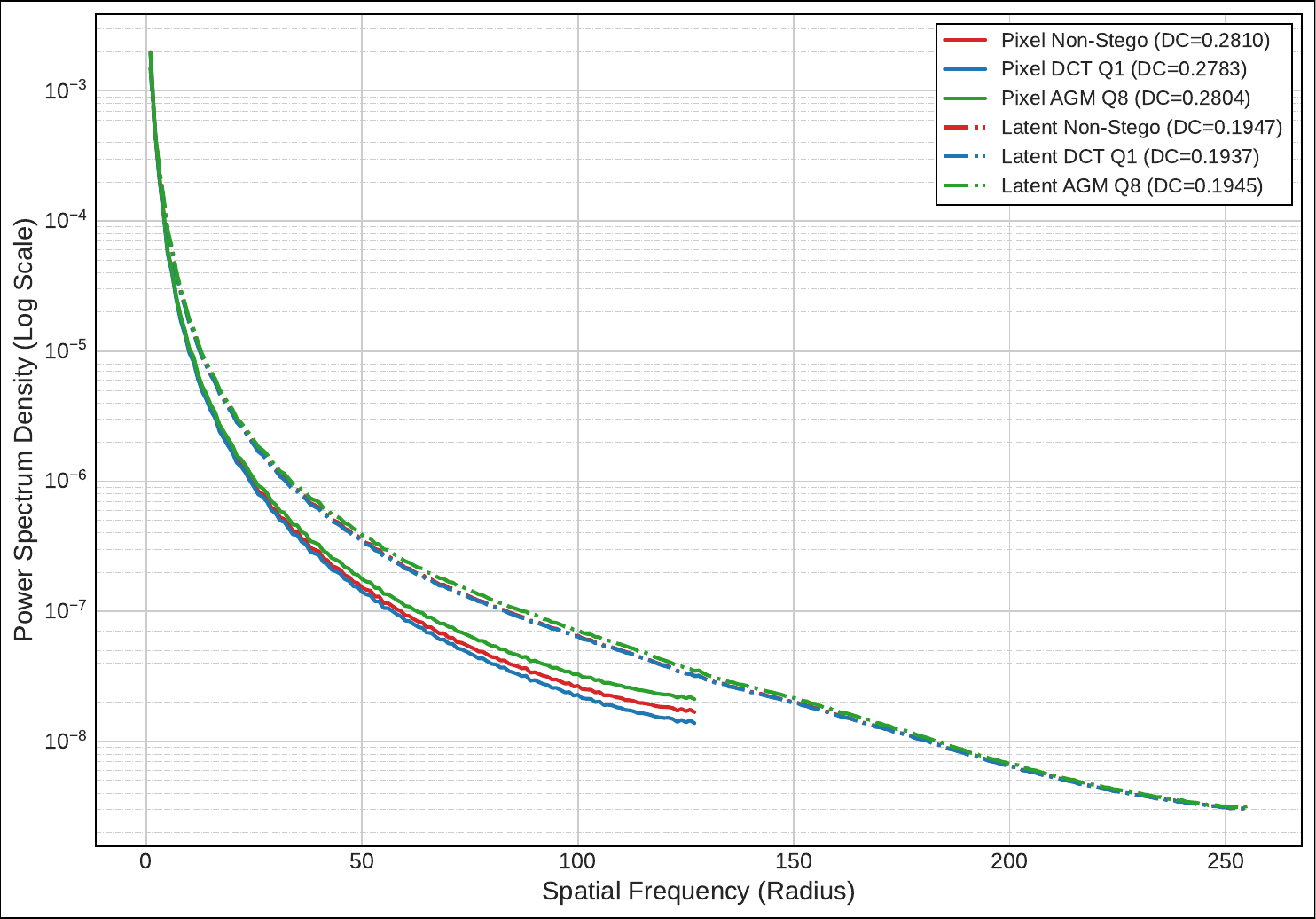}
	\vspace{-1.5em}
    \caption{Average radial Power Spectrum Density in log scale. The spectra are computed over 20,000 samples (normalized from $[0,255]$ to $[0,1]$) for the Non-Stego, AGM ($Q=8$), and IDCT ($Q=1$) sets in both pixel and latent spaces. 
	DC values denote the zero-frequency amplitude (global luminance), where $0.0$ corresponds to absolute black and $1.0$ to white saturation.}
    \label{fig:spectra}
	\vspace{-1.5em} 
\end{figure}

To resolve these interrelated hypotheses, we examine the radial power spectral characteristics visualized in Fig. \cref{fig:spectra}.
In the pixel domain, IDCT ($Q=1$) exhibits a discernible spectral separation from the Non-Stego baseline across medium-to-high frequencies, 
empirically verifying the existence of characteristic spectral artifacts induced by frequency-domain operations. In contrast, the spectral behavior in the latent space demonstrates a significant smoothing effect. 
Critically, this attenuation occurs regardless of whether the noise mapping method operates in the frequency domain. 
AGM ($Q=8$), which performs no frequency transformations, nonetheless introduces high-frequency artifacts in pixel space that are similarly suppressed in latent space. 

Although Fig. \cref{fig:spectra} illustrates a spectral deviation between the latent AGM and Non-Stego baselines, this discrepancy does not translate to effective detection performance. 
Steganalyzers establish decision boundaries based on the statistical distributions of cover and stego samples. 
In the context of latent-space diffusion models, the cover distribution exhibits substantial intrinsic variance driven by semantic diversity. 
The magnitude of spectral perturbations induced by latent embedding remains statistically subordinate to these natural fluctuations distinct to varying text prompts. 
Consequently, the classifier fails to isolate the steganography signal from the inherent variability of the generative process. 

To validate whether semantic conditioning masks artifacts, we isolate the impact of the text encoder by constraining the generation process to a fixed text prompt. 
This control removes the intrinsic variance associated with diverse text prompts and forces the steganalyzer to discriminate based solely on embedding residues. 
\cref{tab:fixed_prompt_ablation} presents the detection error rates $P_E$ under this controlled condition.

The performance degradation relative to \cref{tab:steganalysis} confirms that prompt-induced variance effectively obscures noise mapping artifacts. 
The significant drop in $P_E$ indicates that steganalyzers struggle to differentiate between semantic fluctuations and steganography perturbations in diverse settings. 
Suppressing semantic noise renders the embedding signal statistically distinct.

This exposure is particularly acute for AGM ($Q=8$), where $P_E$ collapses to 0.0023. 
Such vulnerability implies that the distributional shift driven by the large KL divergence is statistically conspicuous but remains submerged within the generative variance. 
The comparative resilience of IDCT ($Q=1$) suggests that AGM induces more severe statistical anomalies that strictly necessitate semantic diversity for concealment.
\begin{table}[!t]
	\centering
	\caption{Latent-Space Steganalysis Performance ($P_E$) Under Fixed Text Prompt Condition.}
	\label{tab:fixed_prompt_ablation}
	
	\scriptsize
	\setlength{\tabcolsep}{3pt}       
	
	\newcolumntype{Y}{>{\centering\arraybackslash}X}
	
	\begin{tabularx}{\columnwidth}{@{} l | Y Y Y Y @{}}
		\toprule
		
		\diagbox[width=7em, trim=l]{Method}{Detector} 
		& SRNet \cite{SRNet} & XuNet \cite{XuNet} & YeNet \cite{YeNet} & SiaStegNet \cite{SiaStegNet} \\
		\midrule
		
		AGM ($Q=8$) & 
		0.0023 & 0.2150 & 0.3941 & 0.0047 \\
		
		IDCT ($Q=1$) & 
		0.4387 & 0.4180 & 0.4752 & 0.4553 \\
		
		\bottomrule
	\end{tabularx}
\end{table}

\paragraph{Structural Regularization of VAE Encoder}
In pixel-space models, the inverse integration operates directly on the raw pixel domain. 
Given the extreme sensitivity of ODE inversion to terminal state deviations, even minor spatial perturbations propagate exponentially, rendering pixel-based steganography highly fragile.
In contrast, latent diffusion models mandate a projection via the VAE encoder $\mathcal{E}(\cdot)$ prior to inverse integration. 

We hypothesize that the encoder functions as a structural regularizer, attenuating the energy of input perturbations. 
To validate this mechanism, we evaluate the relative perturbation amplification ($R$) under varying AWGN intensities $\alpha$ as a case study. 
To ensure a rigorous comparison across domains with distinct dynamic ranges and dimensionalities, we adopt a two-stage normalization process.

First, we define the relative perturbation magnitude, denoted as $\eta(\alpha)$, which normalizes the distortion against the signal energy within its respective domain. 
Let $\mathbf{x}$ denote the clean reference image and $\mathbf{z} = \mathcal{E}(\mathbf{x})$ denote its latent variable. 
We generate the perturbed image $\mathbf{x}'_\alpha = \mathbf{x} + \alpha \cdot \boldsymbol{\epsilon}$ by introducing standard Gaussian noise $\boldsymbol{\epsilon} \sim \mathcal{N}(\mathbf{0}, \mathbf{I})$ scaled by a scalar intensity factor $\alpha$. 
In this evaluation, $\alpha$ is incremented linearly from $0$ to $0.1$ with a step size of $0.001$.
The corresponding perturbed latent representation is obtained as $\mathbf{z}'_\alpha = \mathcal{E}(\mathbf{x}'_\alpha)$. 
The relative magnitudes are computed as:
\begin{equation}
    \eta_{pix}(\alpha) = \frac{\|\mathbf{x}'_\alpha - \mathbf{x}\|_2}{\|\mathbf{x}\|_2}, \quad 
    \eta_{lat}(\alpha) = \frac{\|\mathbf{z}'_\alpha - \mathbf{z}\|_2}{\|\mathbf{z}\|_2}.
\end{equation}

Second, to quantify the rate of error growth independent of initial sensitivity, we normalize $\eta(\alpha)$ against the fundamental perturbation unit $\alpha_0=0.001$. 
The final amplification factor $R$ is formulated as:
\begin{equation}
    R_{pix}(\alpha) = \frac{\eta_{pix}(\alpha)}{\eta_{pix}(\alpha_0)}, \quad 
    R_{lat}(\alpha) = \frac{\eta_{lat}(\alpha)}{\eta_{lat}(\alpha_0)}.
\end{equation}

This metric isolates the perturbation response characteristic, allowing for a direct comparison of structural stability between the pixel and latent spaces.

\begin{figure}[!t]
    \centering
    \includegraphics[width=\linewidth]{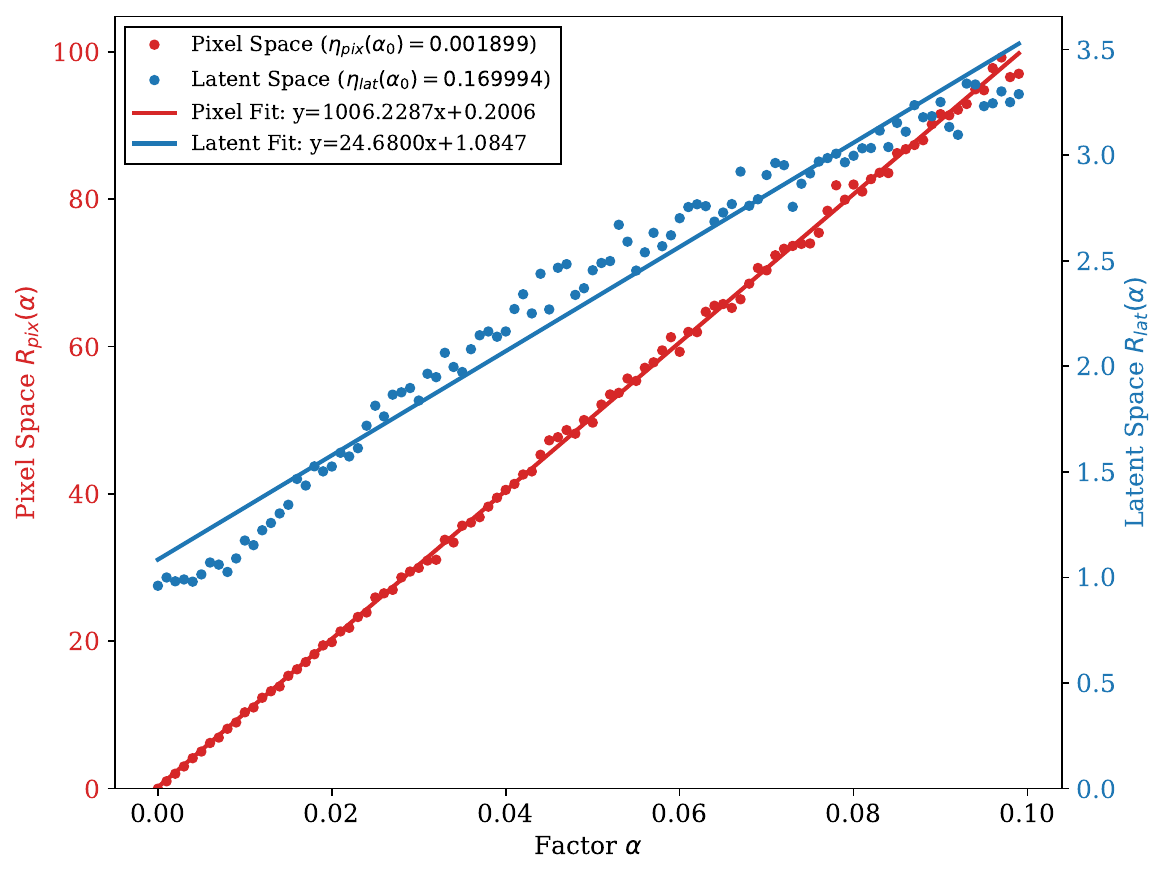}
	\vspace{-2em}
    \caption{Comparative regression analysis of relative perturbation amplification $R$ under AWGN. The significantly lower error growth rate in the latent domain illustrates the structural damping effect of the VAE encoder.}
    \label{fig:vae_robustness}	
\end{figure}

Figure \cref{fig:vae_robustness} presents the comparative regression analysis. 
The pixel-space response $R_{pix}(\alpha)$ exhibits an aggressive linear growth with a slope of $k_{pix} \approx 1006.22$. 
This confirms that without an intermediate buffer, noise energy accumulates directly, leading to catastrophic trajectory deviation in pixel-space inversion. 
Conversely, the latent-space response $R_{lat}(\alpha)$ demonstrates a significantly attenuated growth rate with a slope of $k_{lat} = 24.68$. 
This disparity reveals a damping ratio of approximately $40:1$. 
The VAE encoder effectively suppresses the transmission of stochastic perturbations into the latent representation. 
Consequently, the input to the inverse ODE remains relatively stable despite significant pixel-level corruption. 
This structural characteristic partially accounts for the superior robustness of latent-space methods observed in \cref{tab:robustness}.

\section{Conclusion}
\label{sec:conclusion}
Driven by the empirical finding that controlled deviations from the standard normal prior enhance numerical reversibility, 
this work proposes the Approximate Gaussian Mapping to adaptively negotiate the trade-off between retrieval secret bitstream accuracy and security under fixed capacity constraints.
Comparative evaluation establishes a deployment criterion where pixel-space models maximize embedding capacity and BAR suitable for secure lossless channels, 
whereas latent-space frameworks leverage the structural regularization of the variational autoencoder and semantic masking of CLIP models to deliver the superior robustness and security required for adversarial open environments.
These findings motivate future research to expand beyond the pursuit of statistical alignment and investigate semantic-aware steganography, 
where the intrinsic diversity of the generative process is actively exploited to obfuscate secrets hiding traces while ensuring both high security and retrieval accuracy.

{\appendix[]
\label{app:kl_derivation}
\setcounter{equation}{0}
\renewcommand{\theequation}{A\arabic{equation}}

This appendix derives the analytical approximation for the Kullback-Leibler (KL) divergence between the distribution of the mapped noise variable and the standard normal prior (SNP). 

\subsection{Decomposition of Total Divergence}
Let $\mathbf{x}_T \in \mathbb{R}^N$ denote the secret message bitstream mapped noise vector at the terminal timestep $T$, and let $\mathbf{n} \sim \mathcal{N}(\mathbf{0}, \mathbf{I}_N)$ be the target SNP variable. 
The global objective is to minimize the divergence $D_{KL}(P(\mathbf{x}_T) \| \Phi(\mathbf{n}))$, where $P$ and $\Phi$ denote the multivariate probability density functions (PDFs) of $\mathbf{x}_T$ and $\mathbf{n}$, respectively.

We adopt the widely accepted Independent and Identically Distributed (I.I.D.) assumption. 
The embedding process applies identical statistical operations to each element of the noise vector. Consequently, the joint PDF $P(\mathbf{x}_T)$ factorizes into the product of marginal PDFs $p(x_i)$ for each scalar component $x_i$:
\begin{equation}
    P(\mathbf{x}_T) = \prod_{i=1}^{N} p(x_i), \quad \Phi(\mathbf{n}) = \prod_{i=1}^{N} \phi(n_i),
\end{equation}
where $\phi(\cdot)$ is the standard normal PDF. Utilizing the additivity property of KL divergence for independent distributions, the total divergence decomposes as:
\begin{align}
    D_{KL}(P \| \Phi) &= \int_{\mathbb{R}^N} P(\mathbf{x}_T) \log \frac{P(\mathbf{x}_T)}{\Phi(\mathbf{x}_T)} d\mathbf{x}_T \nonumber \\
    &= \sum_{i=1}^{N} \int_{\mathbb{R}} p(x_i) \log \frac{p(x_i)}{\phi(x_i)} dx_i \nonumber \\
    &= N \cdot D_{KL}(p(x) \| \phi(x)).
\end{align}

This linear relationship implies that minimizing the KL divergence for a single scalar variable $x$ equivalently minimizes the global divergence for the entire vector $\mathbf{x}_T$.

\subsection{Analytical Approximation for Scalar Variable}
We now derive the approximation for the scalar component $x$, constructed as:
\begin{equation}
    x = \frac{u + n}{\sigma}, \quad u = S \left( \frac{m}{2^Q - 1} - 0.5 \right),
\end{equation}
where $n \sim \mathcal{N}(0, 1)$, $m \sim U\{0, \dots, 2^Q-1\}$, and $S, \sigma$ are optimizable scalars.

The derivation employs the Gram-Charlier Type A series expansion, which approximates a near-Gaussian PDF $p(x)$ using cumulants $\kappa_r$. 
Due to the symmetric distribution of both $u$ and $n$, the variable $x$ is symmetric about zero. 
Consequently, all odd-order cumulants $\kappa_{2r+1}$ vanish. The even-order cumulants are derived using the property $\kappa_r(ax) = a^r \kappa_r(x)$ and the additivity of cumulants for independent variables:

\begin{equation}
    \kappa_2(x) = \frac{\text{Var}(u) + 1}{\sigma^2}, \quad \kappa_{2r}(x) = \frac{\kappa_{2r}(u)}{\sigma^{2r}} \quad (r \ge 2),
\end{equation}
where $\kappa_{2r}(u)$ depends solely on the moments of the discrete uniform distribution.

The PDF $p(x)$ is approximated by expanding around $\phi(x)$ using probabilists' Hermite polynomials $H_r(x)$:
\begin{equation}
    p(x) \approx \phi(x) \left[ 1 + \sum_{j=1}^{k} \frac{\kappa_{2j} - \delta_{1j}}{ (2j)! } H_{2j}(x) \right] \triangleq \phi(x) [1 + \mathcal{H}(x)],
\end{equation}
where $\delta_{1j}$ is the Kronecker delta handling the variance mismatch ($\kappa_2 - 1$), and $\mathcal{H}(x)$ represents the perturbation term. Assuming small perturbations ($|\mathcal{H}(x)| \ll 1$), we apply the second-order Taylor expansion $\log(1+y) \approx y - y^2/2$:
\begin{align}
    D_{KL} &\approx \int \phi(x) [1 + \mathcal{H}(x)] \left( \mathcal{H}(x) - \frac{\mathcal{H}(x)^2}{2} \right) dx \nonumber \\
    &\approx \int \phi(x) \left( \mathcal{H}(x) + \frac{\mathcal{H}(x)^2}{2} \right) dx,
\end{align}
where cubic terms are neglected. By the orthogonality of Hermite polynomials, $\int \phi(x) H_r(x) dx = 0$ for $r \ge 1$, the linear term $\int \phi(x) \mathcal{H}(x) dx$ vanishes. 

The remaining quadratic term utilizes the orthogonality property $\int \phi(x) H_m(x) H_n(x) dx = m! \delta_{mn}$. This eliminates cross-terms in $\mathcal{H}(x)^2$, yielding:
\begin{equation}
    \int \phi(x) \mathcal{H}(x)^2 dx = \frac{(\kappa_2-1)^2}{2!} + \sum_{j=2}^{k} \frac{\kappa_{2j}^2}{(2j)!}.
\end{equation}

Substituting this into the integral, we obtain the final differentiable objective function, truncated at the 10th order:
\begin{equation}
    D_{KL} \approx \frac{1}{2} \left[ \frac{(\kappa_2(x)-1)^2}{2} + \sum_{j=2}^{5} \frac{\kappa_{2j}(x)^2}{(2j)!} \right].
\end{equation}

Valid under the regime of small perturbations ($|\mathcal{H}(x)| \ll 1$), this analytical derivation provides an efficient proxy for the true KL divergence, 
eliminating the intractability associated with empirical PDF estimation. 
}

\vfill

\end{document}